\documentclass[usegraphicx,usedcolumn,usenatbib]{mn2e}
\usepackage{graphicx}
\usepackage{txfonts}
\usepackage[T1]{fontenc}
\usepackage{rotating}
\usepackage{accents}
\usepackage{aecompl} 

\newcommand{\Msun}{M$_{\odot}$}

\bibpunct[,]{(}{)}{;}{a}{,}{,}

\begin{document}

\title[The role of molecular gas formation on chemical evolution]
{Galaxy chemical evolution models: The role of molecular gas formation}

\author[Moll\'{a} et al. ]
{Mercedes ~Moll\'{a}$^{1}$\thanks{E-mail:mercedes.molla@ciemat.es},
{\'A}ngeles I. D{\'\i}az$^{2,3}$, Yago Ascasibar$^{2,3}$, and Brad K. Gibson$^{4}$ \\
$^{1}$ Departamento de Investigaci\'{o}n B\'{a}sica, CIEMAT, 28040, Madrid. Spain\\ 
$^{2}$ Universidad Aut\'{o}noma de Madrid, 28049, Madrid, Spain \\
$^{3}$ Astro-UAM, Unidad Asociada CSIC, Universidad Aut{\'o}noma de Madrid,28049, Madrid, Spain\\
$^{4}$ E.A. Milne Centre for Astrophysics, University of Hull, Hull, HU6~7RX, United Kingdom}

\date{Accepted Received ; in original form }

\pagerange{\pageref{firstpage}--\pageref{lastpage}} \pubyear{2011}

\maketitle \label{firstpage}

\begin{abstract}
In our grid of multiphase chemical evolution models \citep{md05}, star formation in the disk occurs in two steps: first, molecular gas forms, and then stars are created by cloud-cloud collisions or interactions of massive stars with the surrounding molecular clouds.
The formation of both molecular clouds and stars are treated through the use of free parameters we refer to as efficiencies.
In this work we modify the formation of molecular clouds based on several new prescriptions existing in
the literature, and we compare the results obtained for a chemical evolution model of the Milky Way Galaxy regarding the evolution of the Solar region, the radial structure of the Galactic disk, and the ratio between the diffuse and molecular components, HI/H$_{2}$.
Our results show that the six prescriptions we have tested reproduce fairly consistent most of the observed trends, differing mostly in their predictions for the (poorly-constrained) outskirts of the Milky Way and the evolution in time of its radial structure.
Among them, the model proposed by \citet{asc17}, where the conversion of diffuse gas into molecular clouds depends on the local stellar and gas densities as well as on the gas metallicity, seems to provide the best overall match to the observed data.
\end{abstract}


\section{Introduction}

Chemical elements appear in the Universe as a consequence of three main processes of production: the Big-Bang nucleosynthesis, the spallation process and the stellar nucleosynthesis. The first one only explains H and He and some traces of other light elements, while the second one transforms, by the collision of cosmic rays, some chemical elements in other lighter by the fragmentation of the weightier ones. The third one is actually the most important in forming the majority of elements of the Universe.

Chemical evolution models are the classical tool to interpret observed elemental abundances, and associated quantities such as gas and stellar surface densities, star formation histories, and the distribution of stellar ages.  Elemental patterns carry the fingerprint of star formation timescales from their birth location, regardless of a star's present-day position.  Chemical evolution codes solve a system of first order integro-differential equations, assuming an analytical star formation (SF) law, an initial mass function (IMF), stellar lifetimes, and nucleosynthetic yields.

In \citet[ hereafter MD05]{md05} we calculated a grid of 440 theoretical galaxy models, applied to 44 different radial distributions of galaxy mass and with 10 different values between 0 and 1 for the molecular gas and star formation efficiencies, which implies 10 different evolutionary histories for each one of the different sizes/masses of galaxies. This grid was calibrated on the Milky Way Galaxy (MWG), by using the data of surface densities of HI, H$_{2}$, stars and SF rate (SFR), and also of C, N, and O abundances, and with time evolution data for the solar region such as the metal-enrichment or the SFR histories. Our model has the advantage over similar ones in the literature of  assuming that stars form in two steps: first the molecular clouds form from the diffuse gas following a classical Schmidt law, then cloud-cloud collisions lead to the formation of stars. Therefore, we derive radial distributions for {\bf both} gas phases in a galaxy. This allows a certain delay in the star formation, doing smoother star formation histories. Results from that work show that, as expected, the SFR radial distributions follow those of the molecular gas; however both showing a drop in the inner regions of the disks, at variance with observations. 

In order to explore those differences further, we have started the computation of a new grid of models including updates to all the relevant ingredients as the stellar yields in \citet{molla15}, or the infall rate law in \citet{molla16}. In the present work, our aim is to improve the predicted $\rm H_{2}$ and SFR radial profiles, while maintaining abundance radial gradients in agreement with those observed.  For that we will study the changes in the molecular gas radial distributions when using different prescriptions existing in the literature about the formation of molecular clouds from the diffuse gas and select which of them is the best one to reproduce the existing observations on the MWG taken as the calibration galaxy.

Section 2 of this work presents some of these new prescriptions taken from \citet{bli06,kru08,kru09,gne11,asc17}, describing the different models we compute. In Section 3 we show our model results for the MWG, comparing them with recent observational data as given in \citet[][ Appendix]{molla15}. In Section 4 we check if our results reproduce the universal radial profile H$_{2}$/HI given by \citet{bigiel08}, analyzing the dependencies of this ratio on the galaxy morphological type and on the dynamical mass. Finally, in Section 5 we draw our Conclusions.

\section{Chemical evolution models: HI to H$_{2}$ conversion prescriptions}

In this section we are going to summarize our basic multiphase chemical evolution model (hereinafter {\sc MULCHEM}) and describe the different prescriptions used in this work, to be included in it, to form molecular clouds, H$_{2}$, from the diffuse gas, HI. 

\subsection{Description of the basic model}

As described in \citet{molla16}, we start computing the radial mass distributions for 16 theoretical galaxies following \citet{sal07} equations, defined in terms of the virial mass of the dark matter haloes, $M_{vir}$, and their associated rotation curves for the galaxy disk and halo.  The virial masses are in the range $M_{\textrm{\scriptsize vir}} \in [5\times 10^{10} - 10^{13}]$\,\Msun, with associated barionic disk masses in the range $M_{\textrm{\scriptsize D}} \in [1.25\times 10^{8} -5.3\times 10^{11}]$\,\Msun. The initial gas in each model collapses onto the disk on timescales based in the \citet{sha06} relationship, which gives the ratio between the galaxy disk and the dark matter halo masses, $M_{\textrm{\scriptsize D}}/M_{\textrm{\scriptsize vir}}$. From the rotation curves, we calculate the radial distribution of the galaxy dynamical mass, $M_{\textrm{\scriptsize tot}}(R)$ as well as the predicted disk mass, $M_{\textrm{\scriptsize D}}(R)$, at the present time. Thus, we obtain the infall rate necessary to produce, at the end of a given model evolution, the appropriate disk for each galaxy dynamical mass.  These infall rates are defined by a collapse time scale $\tau(R)$ given by these two accumulated masses distributions, $M_{\textrm{\scriptsize tot}}(R)$ an$M_{\textrm{\scriptsize D}}(R)$ through the corresponding  mass in each radial region, which we define as 1 kpc wide:

\begin{equation}
\Delta M_{\textrm{\scriptsize tot}}(R)=M_{\textrm{\scriptsize tot}}(<R)-M_{\textrm{\scriptsize tot}}(<R-1)\\\
\end{equation}
\begin{equation}
\Delta M_{\textrm{\scriptsize D}}(R)=M_{\textrm{\scriptsize disk}}(<R)-M_{\textrm{\scriptsize disk}}(<R-1)
\end{equation}

\begin{equation}
\tau(R)=-\frac{13.2}{\ln{\left(1-\frac{\Delta M_{\textrm{\scriptsize D}}(R)}{\Delta M_{\textrm{\scriptsize tot}}(R)}\right)}}\,[\mbox{Gyr}]
\label{eq_infall_rate}
\end{equation}

The infall rates inferred from expression~(\ref{eq_infall_rate}) vary among different galaxies and among the different radial regions within them, and they imply a modest evolution with time of the infall rate for the disks but a strong evolution for bulges.
The radial regions in a disk for a galaxy with  $M_{\textrm{\scriptsize vir}}\sim10^{12}$\,\Msun  (i.e., a MWG-like analog) show  present day infall rates of 
$\dot{M}\sim 0.5$\,\Msun\,yr$^{-1}$ for galactocentric radii $R<13$\,kpc, in agreement with \cite{san08}'s data, being much lower for the outer regions, $R> 13$\,kpc, \citep[see][ for a detailed discussion]{molla16}. 

Once defined our initial scenario with these radial distributions of masses, it is necessary to solve the equation system which  solves
the evolution along the time of each radial region located at a galactocentric distance R, in the halo and in the disk.  In {\sc MULCHEM} we solve the following system:
\begin{small}
\begin{eqnarray}
\frac{dg_{H}(R)}{dt}&=&-\left(\kappa_{h,1}(R)+\kappa_{h,2}(R)\right)g^{n}_{H}(R)-f(R)g_{H}(R)+W_{H}(R)\\
\frac{ds_{1,H}(R)}{dt}&=&\kappa_{h,1}(R)g^{n}_{H}(R)-D_{1,H}(R)\\
\frac{ds_{2,H}(R)}{dt}&=&\kappa_{h,2}g^{n}_{H}(R)-D_{2,H}(R)\\
\nonumber
\frac{dg_{D}(R)}{dt}&=&-\kappa_{c}(R) g^{n}_{D}(R)+\kappa_{a}'(R)c(R)s_{2,D}(R)+\kappa_{s}'(R)c^{2}(R)\\
& & +f(R)g_{H}(R)+W_{D}(R)\\
\frac{dc(R)}{dt}&=&\kappa_{c}(R)g^{n}_{D}(R)-\left(\kappa_{a,1}(R)+\kappa_{a,2}(R)+\kappa_{a}'(R) \right)c(R)s_{2,D}(R)\\
& & -\left(\kappa_{s,1}(R)+\kappa_{s,2}(R)+\kappa_{s}'(R)\right)c^{2}(R)\\
\frac{ds_{1,D}(R)}{dt}&=&\kappa_{s,1}(R)c^{2}(R)+\kappa_{a,1}(R)c(R)s_{2,D}(R)-D_{1,D}(R)\\
\frac{ds_{2,D}(R)}{dt}&=&\kappa_{s,2}(R)c^{2}(R)+\kappa_{a,2}(R)c(R)s_{2,D}(R)-D_{2,D}(R)\\
\frac{dr_{H}(R)}{dt}&=&D_{1,H}(R)+D_{2,H}(R)-W_{H}(R)\\
\frac{dr_{D}(R)}{dt}&=&D_{1,D}(R)+D_{2,D}(R)-W_{D}(R)
\end{eqnarray}
\end{small}
\normalsize

These equations predict the time evolution of the different phases of the model. 
More importantly, we consider in our model MULCHEM \citep{fer92,fer94,md05} two phases of gas in the disk:  diffuse gas, $g$,  and molecular gas, $c$, which are well known to be essential ingredients in the process of star formation, but have only recently been explicitly included in other chemical evolution models \citep[e.g.][]{kub15a,kub15b}. 
{\sc MULCHEM} divides stars in two ranges, $s_{1}$, and $s_{2}$, denoting low-mass and intermediate, and massive stars, respectively, and stellar remnants, $r$. In all cases
 subscripts $D$ and $H$ correspond to disk and halo, respectively.  The infall rate, $f(R)$, is the inverse of the collapse time $\tau(R)$, given by the expression \ref{eq_infall_rate},
and the death rates are computed with:
\begin{eqnarray}
D_{1,H,D}(R,t)&=&\int_{m_{min}}^{m_{*}}\Psi_{H,D}(R,t-\tau_{m})m\phi(m)dm\\
D_{2,H,D}(R,t)&=&\int_{m_{*}}^{m_{max}}\Psi_{H,D}(R,t-\tau_{m})m\phi(m)dm,
\end{eqnarray}
being $m_{*}=4 M_{\sun}$ the mass limiting the low mass stars  from intermediate mass and massive stars, 
 $m_{min}$ and $m_{max}$ the lower and upper mass limits of the initial mass function 
$\phi(m)$, and $\tau_{m}$ the main sequence lifetime of a star of mass $m$.

As explained in MD05, in our model it is assumed that the star formation takes place following a Schmidt law in the halo regions; in the disk, however, it occurs in two steps: first molecular clouds, $c_{D}$, form from diffuse gas, $g_{D}$, then stars form through cloud-cloud collisions. Besides, a second star formation process appears resulting from the interaction of massive stars, $s_{D,2}$  with the molecular clouds surrounding them. Therefore,  we have different processes defined in the galaxy:
\begin{enumerate}
\item Star formation by spontaneous fragmentation of gas in the halo:
$\propto \kappa_{h,1;2}\,g_{H}^{n}$, where we use $n = 1.5$
 \item Disk formation by gas accretion from the halo or protogalaxy: $fg_{H}$
 \item Clouds formation by diffuse gas. In our standard models this process is $\propto \kappa_{c} g_{D}^{n}$.
  (with $ n=1.5$).  This cloud formation law will be modified in the present models as we will describe in next subsections
 \item Star formation due to cloud collision: $\propto \kappa_{s,1;2}c^{2}$
 \item Diffuse gas restitution due to cloud-cloud collision: 
$\propto \kappa_{s}'c^{2}$
 \item Induced star formation due to the interaction between clouds
and massive stars: $\propto \kappa_{a,1;2}c\,s_{2,D}$
 \item Diffuse gas restitution due to the induced star formation \footnote{ Massive stars induce SF in the surrounding
 molecular clouds but their radiation also destroys a proportion of them.}: $\kappa_{a}'c\,s_{2,D}$,
\end{enumerate}
where $\kappa_{h}$, $\kappa_{c}$, $\kappa_{s}$ and $\kappa_{a}$ are the proportionality factors of
the star formation in the halo, the cloud formation, the cloud-cloud collision and the cloud-massive stars
interactions (last two create stars from molecular clouds)\footnote{For the sake of avoiding misinterpretations with other quantities, we have changed the letters $K$, $\mu$, $H$ and
 $a$,  used in our previous works to denote the parameters defining the star formation in the halo, the molecular clouds formation in the disk,  the cloud-cloud collisions and the cloud-massive stars interactions, by  $\kappa_{h}$, $\kappa_{c}$, $\kappa_{s}$, and $\kappa_{a}$, respectively. Thus, the old efficiencies $\epsilon_{K}$, $\epsilon_{\mu}$, $\epsilon_{H}$ and $\epsilon_{a}$  are called now $\epsilon_{h}$, $\epsilon_{c}$, $\epsilon_{s}$, and $\epsilon_{a}$.}. 
Since stars are divided in two groups, $s_{1}$, and $s_{2}$, the parameters involving star formation 
are divided in the two groups too, thus: $\kappa_{h}=\kappa_{h,1}+\kappa_{h,2}$, $\kappa_{s}=\kappa_{s,1}+\kappa_{s,2}+\kappa_{s}' $, and $\kappa_{a}=\kappa_{a,1}+\kappa_{a,2}+\kappa_{a}'$, where  terms $\kappa_{s}'$ and $\kappa_{a}'$ refer to the restitution of diffuse gas due to the cloud-cloud collisions and massive stars-cloud interaction processes.

Thus, the star formation law in halo and disk is: 
 \begin{eqnarray}
 \Psi_{H}(R,t)& =& (\kappa_{h,1}(R)+\kappa_{h,2}(R))g_{H}(R)^{n}\\
 \Psi_{D}(R,t) &=& (\kappa_{s,1}(R)+\kappa_{s,2}(R))c(R)^{2}+(\kappa_{a,1}+\kappa_{a,2})c(R)\,s_{2,D}(R)
\end{eqnarray}
 
The factors $\kappa_{h}$, $\kappa_{s}$,  and $\kappa_{a}$ have  a radial dependence, as discussed in previous studies \citep{fer94,molla14}:

\begin{eqnarray}
\label{par}
\kappa_{h}(R)&=&\epsilon_{h}(G/V_{H}(R))^{1/2}\\
\kappa_{s}(R)&=&\epsilon_{s}(3./V_{D}(R)) \\
\kappa_{a}(R)&=&\epsilon_{a}(G\rho_{c})^{1/2}/<m_{s_{2}}>,
\end{eqnarray}
where $G$ is the universal gravitational constant, $V_{H}(R)$ and $V_{D}(R)$ are the halo and the disc volumes of each radial region, $\rho_{c}$ is the average cloud density and $<m_{s_{2}}>$ is the average mass of massive stars. Thus, we converted these parameters in relationships with the volume and with other quantities that we called {\sl efficiencies},  
$\epsilon_{h}$,  $\epsilon_{s}$ and $\epsilon_{a}$, which represent probabilities (in the range $[0,1]$) associated to the processes of conversion among the different phases and that we assume constant for all radial regions within a galaxy. The efficiency to form stars in the halo, $\epsilon_{h}$, is obtained through the selection of the best value $\kappa_{h}=9\times10^{-3}$, able  to reproduce the SFR and abundances of the Galactic halo \citep[see][ for details]{fer94}. We assumed that it is approximately  constant for all halos. The last efficiency, $\epsilon_{a}$, was also obtained from the best value $\kappa_{a}=2.5\times10^{-8}$ for the MWG, and it is also assumed constant for all galaxies, since these interactions between clouds and massive stars are local processes.  Thus, we have only one free parameter: $\epsilon_{s}$. 
In the present work we are going to modify the formation of H$_{2}$, keeping this efficiency  $\epsilon_{s}$ as a free parameter with values in the range [0--1] as in MD05 and \citet{molla14}. We compute 10 possible values using:
\begin{equation}
\label{efi1}
\epsilon_{s}=exp^{-NT^{2}/8},
\end{equation}
where NT is a free parameter (although related to the morphological type, see details in Sections 2.2 and 2.3), with 10 assigned values between 1 and 10, as given in column 2 of Table~\ref{efis}.

The equations of the chemical abundances are:
\begin{equation}
\frac{X_{i,H}(R)}{dt}=\frac{W_{i,H}(R)-X_{i,H}(R)W_{H}(R)}{g_{H}(R)}
\end{equation}
\begin{equation}
\frac{X_{i,D}(R)}{dt}=\frac{W_{i,D}(R)-X_{i,D}(R)W_{D}(R)+f(R)g_{H}(R)[X_{i,H}(R)-X_{i,D}(R)]}{g_{D}(R)+c(R)}
\end{equation}
where $X_{i}$ are the mass fractions of the 15 elements considered by the model:
$^{1}$H, D, $^{3}$He, $^{4}$He, $^{12}$C, $^{16}$O, $^{14}$N, $^{13}$C, $^{20}$Ne, $^{24}$Mg, $^{28}$Si, $^{32}$S, $^{40}$Ca,
$^{56}$Fe, and the rich neutron isotopes created from $^{12}$C, $^{16}$O, $^{14}$N and from $^{13}$C; and the restitution rates are:
\begin{equation}
W_{i,H,D}(R,t)=\int_{m_{min}}^{m_{max}} \left[ \sum_{j}{\tilde{Q}}_{ij}(m)X_{j}(t-\tau_{m})\right]\Psi_{H,D}(t-\tau_{m})dm
\end{equation}

To compute the elemental abundances we use the technique based in the {\sl matrices $Q$ formalism} \citep{ta73,fer92,pcb98}. Following this formalism, each
element $(i,j)$ of a matrix, $Q_{i,j}$ gives the proportion of a star which was element $j$ and is ejected as $i$ when the star dies. Thus,
\begin{equation} 
Q_{i,j}(m)=\frac{m_{i,j,exp}}{m_{j}},
\end{equation}
and 
\begin{equation}
Q_{i,j}(m)X_{j}=\frac{m_{i,j,exp}}{m}.
\end{equation}

 If we take into account the number of stars of each mass $m$, given by the initial mass function,  that eject this mass $m_{i,j}$, 
 we have:
 \begin{equation}
 \tilde{Q}_{i,j}(m)=Q_{i,j}m \phi(m),
 \end{equation}
 and, therefore, the term $\sum\limits_{j}{\tilde{Q}}_{i,j}(m)X_{j}$ in Eq. (22) represents the mass of an element $i$ ejected by all stars of mass $m$.  This method allows to relax the hypothesis of solar proportions in the ejection, since each element $i$ relates with its own sources. It was originally introduced to compensate for the lack of stellar models of different metallicities,  when the dependence  on Z was not included in the stellar yields calculations  Now that stellar yields for different metallicities are available, the use of  $Q$ matrices allows to us  to take into account possible differences of chemical composition within a given Z. Stellar yields 
 have usually been computed  assuming only solar relative abundances among the different elements at a given Z. However, the relative abundances of elements are not always solar nor constant along the evolutionary time. 
Therefore, as \cite{pcb98} explained, "{\sl The $Q_{i,j}$ matrix links any ejected species to all its different nucleosynthetic sources, allowing the model to scale the ejecta with respect to the detailed initial composition of the star through the Xj's}".
 
Following \citet{molla15}, we use the stellar yield sets from \citet{lim03,chi04} for massive stars, together with yields from \citet{gav05,gav06} for low and intermediate mass stars, combined with the IMF from \citet{kro01}, which is one of the best combinations able to reproduce the MWG data. For supernovae type Ia (SNe-Ia) we use the rates given by \citet{rlp00}, who provides a numerical table (private communication) with the time evolution of the supernova rates for a single stellar population, computed under updated assumptions about different scenarios and probabilities of occurrence. The stellar yields for SNe-Ia are those given by \citet{iwa99}. As shown in \citet{molla15}, the observational data for the MWG are well reproduced with our {\sc MULCHEM} model. The set of data used for comparison with the results of our models were included as Appendix A in \citet{molla15} and refers to a) the time evolution of
the Solar Region --located at a galactocentric distance of 8 kpc--, (star formation and enrichment histories, as $\Psi(t)$ and $[\rm Fe/H](t)$, alpha-element over-abundances 
$[\rm \alpha/Fe]$, as a function of the metallicity $[\rm Fe/H]$, and the metallicity distribution function, MDF, or proportion of stars in each [Fe/H] bin);  b) the radial distributions within the Galactic disk for the present time ($HI$ and $H_{2}$ surface densities, $\Sigma_{HI}$ and $\Sigma_{H_{2}}$, the stellar profile $\Sigma_{\star}$, and the radial distribution of the star formation rate surface density, $\Sigma_{SFR}$); and c) the radial gradients of C, N and O elemental abundances.

\subsection{Our standard model}

The molecular cloud phase in our standard model follows the equation:
\begin{equation}
\frac{dc(R)}{dt}= \kappa_{c}(R)\, g_{D}(R)^{n}-\kappa_{s}(R)\,c^2(R)-\kappa_{a}\,c(R)\,s_{2,D}(R),
\end{equation}
where the first term refers to the formation of molecular cloud from diffuse gas, while the next two others describes how these clouds disappear
due to the cloud-cloud collisions (which create stars and also restitute diffuse gas to the ISM) and as the consequence of the interaction of massive stars
with the cloud surrounding them (this process also create stars and new diffuse gas which return to the ISM).

The proportionality factor  $\kappa_{c}$ depend on the geometry of the regions, in a similar way that the explained for the three others in Eq.~\ref{par},  20, and 21:
\begin{equation}
\kappa_{c}(R)=\epsilon_{c}(G/V_{D}(R))^{1/2}\\
\label{kc}
\end{equation}

In  MD05 we used the data from \citet{young96} corresponding to  $g_{D}$ and $c_{D}$ -- as calculated from the masses of HI and H$_{2}$--, and the SFR, $\Psi_{D}$,  to obtain a relation between both efficiencies, 
$\ln{\epsilon_{c}}$ and $\ln{\epsilon_{s}}$, as a function of  the galaxy morphological type (see Fig.~4 from MD05), finding that   $\ln{\epsilon_{c}}/\ln{\epsilon_{s}}\sim 0.34$. Using Eq.~\ref{efi1} for 
$\epsilon_{s}$, it implies:
\begin{equation}
\epsilon_{c}=exp^{-NT^{2}/20}
\label{old}
\end{equation}

We computed this way both efficiencies simultaneously as a function of the free parameter $NT$, that may be associated to the morphological type index (see MD05 for details about this relationship). The resulting efficiencies $\epsilon_{c}$ are given in column 3 of Table~\ref{efis}. The model using these prescriptions is called STD. 

\subsection{Our modified model}

Looking at Fig.~4 from MD05, it can be actually seen that,  although a constant value for the ratio 
$\ln{\epsilon_{c}}/\ln{\epsilon_{s}}$ is statistically significant, a straight line fit is also reasonable:
\begin{equation}
 \frac{\ln{\epsilon_{c}}}{\ln{\epsilon_{s}}}=0.12+0.07\,NT
\end{equation}
 As for $ \epsilon_{s}$, the efficiency to form stars from molecular clouds, we have kept their values as in MD05, we now modify $\epsilon_{c}$ accordingly: 
 
\begin{equation}
 \epsilon_{c}= exp^{-\frac{NT^{2}}{8}\,(0.12+0.07\,NT)}
\label{new}
\end{equation}

 This changes slightly the values for both efficiencies as shown 
 in Table~\ref{efis} where, for each value of $NT$ (column 1), we give the efficiency to form stars, 
 $\epsilon_{s}$ (column 2), and the old and new values for the molecular gas formation efficiency,
 $\epsilon_{c}$ (columns 3 and 4, respectively).
 
\begin{table}
\caption{Values for the efficiencies of star and molecular gas formation as a function of the parameter (related to morphological type) NT.}
\begin{tabular}{rrrr}
$NT$ & $\epsilon_{s}$ & $\epsilon_{c}$(MD05) & $\epsilon_{c}$\\
\hline
1 & 0.882    &   0.951& 0.977     \\
 2 & 0.607    & 0.819 &  0.878     \\
 3 & 0.325    &   0.638 & 0.699     \\
 4 & 0.135    &   0.449 & 0.449     \\
 5 & 0.044    &   0.287 & 0.230   \\
 6 & 0.011    &   0.165 & 0.088    \\
 7 & 0.002    &   0.086 & 0.023     \\
 8 & 3.35E-4  &   0.040 & 0.004    \\
 9 & 4.00E-5  &   0.017 & 5E-04  \\
10 & 3.70E-6  &   0.007 & 3E-05   \\
\hline
\end{tabular}
\label{efis}
\end{table}

\begin{figure}
\includegraphics[width=0.35\textwidth,angle=-90]{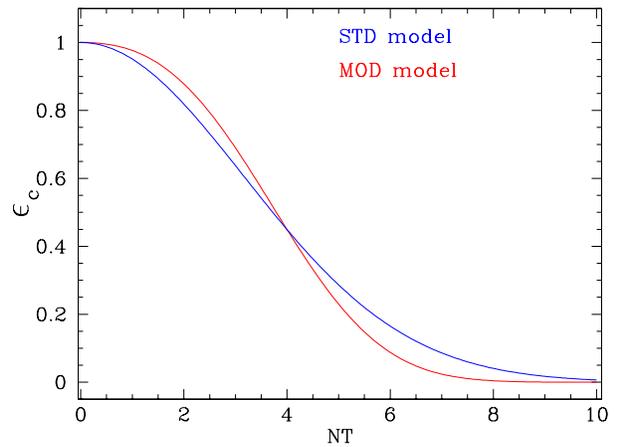}
\caption{ The dependence of the efficiency to form molecular clouds --$\epsilon_{c}$-- on the parameter $NT$ (related to the morphological type), for our STD (Eq. 8) and MOD (Eq.11) models as labelled.}
\label{emu}
\end{figure}

In Fig.~\ref{emu} we plot $\epsilon_{c}$ as a function of $NT$ both for the STD model (Eq. ~\ref{old}) and our modified model (Eq. ~\ref{new}) that we call MOD.  It is evident that according to the newly adopted  function, late type galaxies will form a smaller amount of molecular gas, while this phase will be 
increased for the earlier types, as compared to the STD model. The equation defining the molecular cloud mass, as the corresponding parameter
 $\kappa_{c}$ are the same for both models, STD and MOD. Only the efficiencies  $\epsilon_{c}$ change among them.

\subsection{Ascasibar et al. (2017) prescriptions}

In this case, we follow a prescription based on \citet{asc17} to calculate the value of the parameter $\kappa_{c}(R)$. The most relevant feature of this model is that the time scale for the formation of molecular clouds:
\begin{equation}
\kappa_{\rm c,ASC}(R) = 2\,n_{\rm dust}(R)\,\langle \sigma v \rangle_{\rm dust},
\end{equation}
which depends on the local number density of dust grains $n_{\rm dust}(R)$ and the thermally-averaged cross-section $\langle \sigma v \rangle_{\rm dust}$ for the condensation of hydrogen molecules onto their surface. We assume that the number of dust particles is proportional to the metal content of the gas, i.e. $n_{\rm dust}\propto n_{\rm H}\,Z$, and we set the normalization $\langle \sigma v \rangle_{\odot} = 6\times 10^{-17}$\,cm$^{3}$\,s$^{-1}$  to the reaction rate estimated by \citet{draine96} for a gas of solar composition \citep[$Z_{\odot}=0.014$, ][]{asp09} at a temperature $T=100$\,K. This way:
\begin{equation}
\kappa_{\rm c,ASC}(R) \sim  2\,n_{\rm H}(R)\,\langle \sigma v \rangle_{\odot} \frac{ Z(R)+Z_{\rm eff} }{ Z_{\odot}}
\label{tauc_yago}
\end{equation}
Other physical processes unrelated to the dust grains (e.g. the $H^{-}$ channel) also contribute to the formation of hydrogen molecules. Although such alternative mechanisms are only relevant for the lowest metallicities, they are extremely important in the early universe, and we represent their combined action by an effective term $Z_{\rm eff} = 10^{-3}\, Z_{\odot} = 1.4 \times 10^{-5}$ that becomes dominant when $Z<Z_{\rm eff}$ \citep[see e.g.][]{glo07}.

The gas density $n_{H} = n_{g} + 2n_{c}$ includes the contribution of both diffuse gas and molecular clouds, and, using the classical gas equation, we set each one as:
\begin{equation}
n_{\{\rm g,c\}}(R) \sim \frac{P(R)}{ k_{\rm B}\,( T_{\rm \{g,c\}} + T_{\rm eff} ) }
\label{nh}
\end{equation}
where $k_{\rm B}$ denotes the Boltzmann's constant, and we adopt $T_{\rm g} = T_{\rm c} = 100$\,K.
In addition to thermal pressure, non-thermal pressure support from turbulent motions, cosmic rays, and magnetic fields is accounted for by an extra term with $T_{\rm eff} = 10^4$\,K (implying, under the assumption of equipartition, a velocity dispersion of the order of $\sim 5.2$~km/s for the diffuse gas and $\sim 3.7$\,km/s for the molecular clouds).
$P(R)$ is the mid-plane pressure which depends, in turn, on the total and gas surface densities \citep[see e.g.][]{elm89,leroy08}:
\begin{equation}
P(R) \approx \frac{\pi}{2}\,G\,\Sigma_{\rm gas}(R) \left[ \Sigma_{\rm gas}(R) + \Sigma_{\star}(R) \right]
\end{equation}
Substituting in equation~(\ref{tauc_yago}), one finally arrives to:
\begin{equation}
\kappa_{\rm c,ASC}(R) =2.67\,\Sigma_{\rm gas}(R) \, [\Sigma_{\rm gas}(R) + \Sigma_{\star}(R)]
\left[ Z(R) + 1.4\times 10^{-5} \right]
\label{eq_asc}
\end{equation}
with surface densities in $\rm M_\odot\,pc^{-2}$ units.
The evolution of the molecular mass is given by:
\begin{equation}
\frac{dc(R)}{dt}=\kappa_{\rm c,ASC}(R)g_{D}(R)-\kappa_{s}(R)\,c^2(r)-\kappa_{a}\,c(R)\,s_{2,D}(R),
\end{equation}
in a similar way to models STD and MOD with $n=1$, (of course, it is also necessary to modify this index in the term associated to the formation of molecular clouds in the equation for the diffuse gas).
We will refer to this prescription as ASC.

\begin{figure}
\includegraphics[width=0.45\textwidth,angle=0]{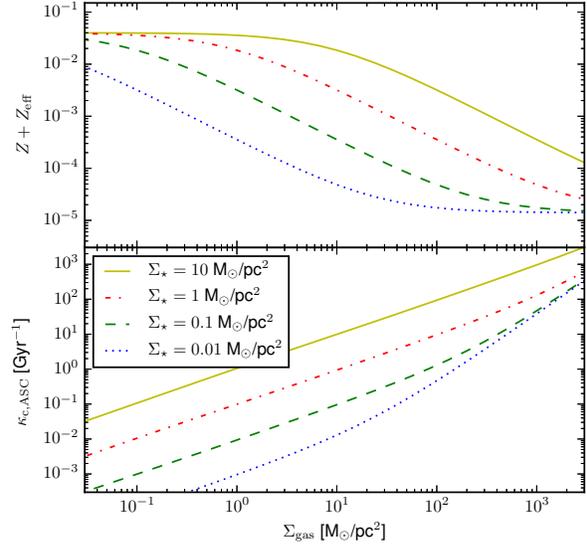}
\caption{Top panel) Rough estimate of the gas metallicity (including an effective term $Z_{\rm eff} = 10^{-3}\, Z_\odot$) according to the expression~(\ref{eq_asc15}); 
Bottom panel) Coefficient $\kappa_c$ of the ASC model, equation~(\ref{eq_asc}), assuming the metallicity plotted on the top panel. In both cases lines represent different stellar surface densities, as indicated in the legend.}
\label{tauc}
\end{figure}

The parameter $\kappa_{\rm c,ASC}$ is represented in Fig.~\ref{tauc} as a function of the total gas (atomic $+$ molecular) surface density for different stellar densities, as labelled. 
As this ratio also depends on the metallicity, we have used the local relation between metallicity and stellar-to-gas fraction proposed by \citet{asc15}, to compute the parameter 
$\kappa_{\rm c,ASC}$ drawn in that Fig.\ref{tauc}:
\begin{equation}
\frac{ Z(S) }{ Z_{\rm max} } = \frac{ S }{ 1 + S/S_{\rm crit} }
\label{eq_asc15}
\end{equation}
where $S(R) \equiv \frac{ \Sigma_{\star}(R) }{ \Sigma_{\rm gas}(R) }$ denotes the surface density ratio between stars and total gas in each radial region, 
 $Z_{\rm max} = 0.032$ is the asymptotic metallicity attained in the {\sl chemically-evolved} limit $S\to\infty$, and $S_{\rm crit}=1.6$ marks the transition between the 
 {\sl chemically-young} regime, where $Z \simeq Z_{\rm max} \frac{S}{ S_{\rm crit} }$, and the flat asymptotic behaviour $Z \simeq Z_{\rm max}$ \citep[see][ for more details]{asc15}. This equation (\ref{eq_asc15}) is, however,  used here only for purposes of doing the figure. In the ASC model the metalllicity is obtained as a result in each radial region and time step. In practice, $ \kappa_{\rm c,ASC}(R) $ is roughly proportional to $ \Sigma_{\rm gas}^2 $ when $ \Sigma_{\rm gas} \gg \Sigma_{\star} $ and $Z \ll Z_{\rm eff}$ (i.e. in the very first stages of galactic evolution) and to the product $ \Sigma_{\rm gas} \Sigma_{\star} $ when $Z \gg Z_{\rm eff}$ (i.e. over most of the life of the galaxy).

\subsection{Gnedin and Kravtsov (2011) prescriptions}

In this case, as in the next two cases, we are going to use the prescriptions given by some authors which calculate the relation between the
molecular gas and the total gas, defined through the ratio:
\begin{equation}
f_{H_{2}}=\frac{\Sigma_{H_{2}}}{\Sigma_{gas}},
\end{equation}
which is equivalent to $c/(g_{D}+c)$ in our notation (since the area is the same for both surface densities) .  This implies that:
\begin{equation}
c(R)=\frac{f_{H_{2}}}{1-f_{H_{2}}}g_{D}(R),
\end{equation}
and consequently,
\begin{equation}
\frac{dc(R)}{dt}=\frac{f_{H_{2}}}{1-f_{H_{2}}}\frac{dg_{D}(R)}{dt}+g_{D}(R)\frac{d}{dt}\left(\frac{f_{H_{2}}}{1-f_{H_{2}}}\right)
\label{derivada}
\end{equation}

\citet[][ hereinafter GNE]{gne11} presented a detailed description of a phenomenological H$_{2}$ formation model.  Their
results give the atomic-to-molecular transition by means of the ratio $f_{H_{2}}$ between the molecular and the total hydrogen, which
they give as a function of gas density or column density with dependencies on the dust-to-gas ratio, $D_{\rm MWG}$ and on the radiation flux at 1000 \AA , $U_{\rm MWG}$. They  parametrize these dependencies in convenient fitting formula useful for their inclusion in semi-analytic models and cosmological simulations that do not include radiative transfer and H$_{2}$ formation. 

\begin{figure}
\includegraphics[width=0.45\textwidth,angle=0]{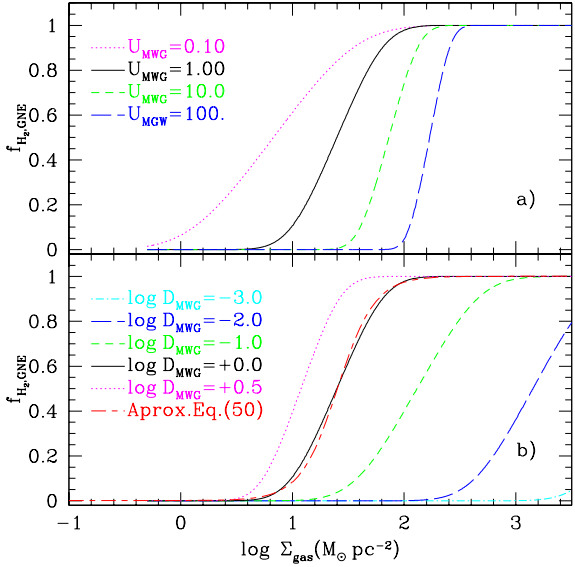}
\caption{The fraction f$_{H_{2}}$ following the prescriptions used in GNE.
a) Dependence of $f_{H_{2}}$ for $D_{\rm MWG}=1.0$  on the parameter U$_{\rm MWG}$ with different values as labelled. 
b) Dependence of $f_{H_{2}}$ for $U_{\rm MWG}=1.0$  on the parameter D$_{\rm MWG}$ with different values as labelled. 
The red line in the bottom panel is the approximated expression as given by Eq.(50) --see text. }
\label{gne}
\end{figure}

Values for $D_{MWG}$ range from  10$^{-3}$ to 3 relative to the MWG value, and  scaling to metallicity, while those for $U_{MWG}$, are between 0.1 and 100 times the MWG value. The ratio $f_{H_{2}}$ is approximated by the equation:
\begin{equation}
f_{H_{2}, GNE}=\frac{1}{1+{\rm exp}(-4x_{GNE}-3x_{GNE}^{3})},
\label{eq_gne}
\end{equation}
where 
\begin{eqnarray}
x_{GNE}& =& \Lambda^{3/7}\,\ln\left({D_{\rm MWG}\frac{n_{H}}{\Lambda n_{*}}}\right),\\ 
n_{\star}&=&25\,cm^{-3},\\
\Lambda&=&\ln{(1+g_{GNE}\,D_{\rm MWG}^{3/7}(U_{\rm MWG}/15)^{4/7})}
\end{eqnarray}
The function $g_{GNE}$ is a function giving the transition between self-shielding and dust shielding, defined as:
\begin{eqnarray}
g_{GNE}&=&\frac{1+\alpha_{GNE}\,s_{GNE}+s_{GNE}^{2}}{1+s_{GNE}}, \\
{\rm with}:\\
s_{GNE}& =& \frac{0.04}{D_{*}+D_{\rm MWG}}, \\
\alpha_{GNE}&=&5\frac{U_{\rm MWG}/2}{1+(U_{\rm MWG}/2)^{2}}\\
D_{*}& = &1.5\times10^{-3}\times \,\ln{(1+(3U_{\rm MWG})^{1.7})}
\end{eqnarray}
In these equations $\rm n_H$ refers to the hydrogen particle total density in their different phases and $\rm n_*$ denotes a fiducial value included only
for normalization purposes in (32). 

We have computed these functions for 4 values of $U_{\rm MWG}=0.1$,1,10 and 100 and 8 values of $D_{\rm MWG}$, which
is given by its relation with the metallicity: $D_{\rm MWG}=\frac{Z}{Z_{\odot}}$, from $10^{-3}$ to $10^{0.5}$.
The results are shown in Fig.~\ref{gne} for a value of $D_{\rm MWG}=1.0$ in the top panel, only modifying $U_{\rm MWG}$,
and for a value of  $U_{\rm MWG}=1$ for variable values of $D_{\rm MWG}$ in the bottom one. 
With these equations, the ratio  $f_{H_{2}}$ depends mainly on the gas surface density with the two other parameters 
modifying the final results: $U_{\rm MWG}$ changing slightly the slope, and $D_{\rm MWG}$ shifting the curve to the right or the left. 
In fact, in STD and MOD models this dependence on the gas surface density also appears, through the disk volume in Eq.~\ref{kc}, which 
finally translates into a radial dependence, {\it i.e.} the diffuse to molecular cloud conversion depends on the gas density in each radial region. In the \citet{gne11} case, it also depends on the metallicity (by means of $D_{\rm MWG}$)  and the parameter $U_{\rm MWG}$.

Such as we may see in that Fig.~\ref{gne}, the function $f_{H_{2}}$ may be approximated as a function $\tanh{(u)}$ with the adequate transformation of the x-axis, using $u=A*(x-B)$ (where $x$ is the $\log{\Sigma_{g_{D}+c}}$). For instance, for $D_{\rm MWG}=1.0$ and $U_{MWG}=1$, 
we find:
\begin{equation}
f_{H_{2}}=\frac{1}{2}{\tanh{[3\times(x-1.7)]}+1}
\label{fit}
\end{equation}
We plot this fitted function in Fig.~\ref{gne}b) as a red short-long-dashed line.

By assuming this dependence we may calculate:
\begin{eqnarray}
\frac{df_{H_{2}}}{dt}&=&\frac{d\tanh{(u)}}{du}\frac{du}{dt}\\
&=& \frac{1}{2}{\rm sech}^2{(u)}\,A\left(\frac{d \log{\Sigma_{g_{D}+c}}}{dt}\right)\\
&=&\frac{A}{2}{\rm sech}^2{u}\frac{1}{g_{D}+c}\left(\frac{dg_{D}}{dt}+\frac{dc}{dt}\right)
\end{eqnarray}

Using this last expression and Eq.\ref{fit} within Eq.\ref{derivada}, and with the adequate mathematics, we obtain
this equation for the formation of molecular clouds from diffuse gas:
\begin{equation}
\frac{dc_{for}}{dt}=\kappa_{c}\frac{dg_{D}}{dt},
\end{equation}
where:
\begin{eqnarray}
\kappa_{c}&=&\frac{f_{H_{2}}+(A/2){\rm sech}^2(u)}{1-f_{H_{2}}-(A/2){\rm sech}^2(u)}\\
& = & \frac{1+\tanh{u}+A{\rm sech}^{2}(u)}{1-\tanh{u}-A{\rm sech}^{2}(u)}
\end{eqnarray}

The function ${\rm sech}(u)$ tends to zero for lower and higher gas densities (see Fig.\ref{gne}), that is, the derivative
of the function $f_{H_{2}}$ when this is almost flat, is practically zero, what it occurs at the beginning and at the end of each curve.
 In this case $\kappa_{c}\sim \frac{f_{H_{2}}}{1-f_{H_{2}}}$. In the middle of the function, when this increases rapidly from 0 to 1,
the function ${\rm sech}(u)$ has a significant value which may reach a value of 1. 
We have estimated the value of $\kappa_{c}$ for the case $U_{MWG}=1$ and $D_{MWG}=1$ for the region 
defined by $ 0.5 \le x \le 3$, or $ -3.5 \le u \le 3.5$.
It results that the expression $1-f_{H_{2}}-(A/2){\rm sech}^2(u)$ began to be negative when $u=-0.6$ or $x=1.5$, that is, $\kappa_{c}<0$. 
We could interpret that the molecular clouds start to be destroyed when $\Sigma_{g+c}\sim 30\,M_{\sun}\,pc^{2}$. Since we
are interested only in the formation processes, we prefer to assume that the formation of molecular clouds is proportional to 
$\sim \frac{f_{H_{2}}}{1-f_{H_{2}}}$ , and computing this destruction with the usual equations used in our code. 
This way, our final equation for the evolution of both phases
of gas in the disk will be given by:

\begin{eqnarray}
\frac{dc(R)}{dt}&=&\kappa_{c}(R)\frac{dg_{D}(R)}{dt}-\kappa_{s}(R)\,c^2(r)-\kappa_{a}\,c(R)\,s_{2,D}(R)\\
\frac{dg_{D}(R)}{dt}&=&\frac{\kappa'_{s}(R)\,c^2(r)+\kappa'_{a}\,c(R)\,s_{2,D}(R)+f(R)g_{H}(R)+W_{D}}{1+\kappa_{c}(R)}
\end{eqnarray}
with 
\begin{equation}
\kappa_{c,GNE}(R)=\frac{f_{H_{2}}}{1-f_{H_{2}}},
\end{equation}
calculated with the expressions given by Eq.~\ref{eq_gne} and the other terms defined as in the STD and MOD models.

In order to include the above equations into our chemical evolution code, we use, as said, the metallicity to calculate
the parameter $D_{\rm MWG}$ at each time, and the total gas density to compute the hydrogen density $n_{H}$. To compute
the $x_{GNE}$ we need to use the density $n_{H}$. We have assumed that this density may be calculated with the surface density of the total gas 
$(g+c)$ dividing by the width of the disk $h\sim 200\,pc$ and doing the adequate change of units from $M_{\sun}\,pc^{3}$ to $cm^{-3}$, 
we obtain that $n_{H}\sim \Sigma_{(g+c)}/5$. Regarding the UV flux, we have assumed $U_{\rm MWG}=1$ for all our simulations since we are modeling MWG. Moreover, it is difficult to evaluate its value in our  chemical evolution models prior to the computation of the photometric evolution of each theoretical galaxy (only masses in the different phases are calculated in this step). In panel a) of Fig.\ref{fh2} we represent the $f_{H_{2}}$ as a function of the surface gas density for different values of the metallicity which, in our code, as also does the gas density, varies continuously when the star formation takes place.  We label models following the prescriptions described in this section, GNE.

\subsection{Blitz \& Rosolowsky (2006) prescriptions}

\citet{bli06} proposed that the ratio of atomic to molecular gas depends basically on the total pressure at each 
radius by the equation:
\begin{equation}
R_{mol}(R)=\frac{M_{\rm{H_{2}(R)}}}{M_{\rm HI}(R)}=\left[\frac{P(R)}{P_{0}}\right]^{\alpha_{P}},
\end{equation}
being $\alpha_{P}$ a parameter.

Using the expression:
\begin{equation}
P(R)=\frac{\pi}{2}G\Sigma_{gas}(R)\left[\Sigma_{gas}(R)+f_{\sigma}\Sigma_{\star}(R)\right],
\end{equation}
where $G$ is the gravitational constant, $R$ is the galactocentric radius and $f_{\sigma}(R) =\sigma_{gas}(R)/\sigma_{\star}(R)$,
is the ratio of the vertical velocity dispersions of gas and stars. Assuming an exponential function for the disk density, and a mean value for $f_{\sigma}(R)$ along the disk, they arrived to the final equation:
\begin{equation}
R_{mol}(R)=1.38\,10^{-3}\left[\Sigma_{gas}(R)+0.1\Sigma_{gas}(R)\sqrt{(\Sigma_{\star}(R)\Sigma_{\star,0}}\right]^{0.92}
\label{eq_bli}
\end{equation}

The equations giving the  time evolution of the diffuse and molecular phases of gas are in this case the same as used in the GNE model. 
Since $R_{mol}=\frac{c}{g_{D}}$ in our notation, we have a similar $\kappa_{c,BLI}=f_{H_{2}}/(1-f_{H_{2}})=R_{mol}$, calculated with
Eq.\ref{eq_bli}, while $f_{H_{2},BLI}=R_{mol}/(1+R_{mol})$.

In this case the molecular ratio depends primarily on the gas surface density but also on the stellar surface density, which may be important when the disk galaxies are gas-poor. At any rate this equation refers to the total pressure averaged over a particular galactocentric distance, which may present problems for its application to cosmological simulations since no correction for clumpliness or the presence of a warm phase are taken into account, but it is totally valid for our purposes.

In panel b) of Fig.\ref{fh2} we represent $f_{H_{2}}$ for different values of the stellar surface density. Obviously in our code this
density, as well as the gas surface density, varies continuously when the star formation takes place. Models calculated following this prescription are labelled BLI.

\subsection{Krumhold et al. (2008, 2009) prescriptions}

\citet{kru08,kru09} studied the transition of the diffuse into molecular gas finding that the conversion depends on the  interstellar radiation field, through
a dimensionless strength $s$, here renamed $s_{KRU}$, which includes the dependence on the dust properties (basically through the metallicity $Z$), the radiation field and the atomic gas density as follows:
\begin{equation}
f_{H_{2}, KRU}=1-\frac{3}{4}\left(\frac{s_{KRU}}{1+0.25s_{KRU}}\right),
\label{eq_kru}
\end{equation}
where 
\begin{equation}
s_{KRU}=\frac{\ln{(1+0.6\chi+0.01\chi^{2})}}{0.04 (Z/Z_{\odot}) \left(\frac{\Sigma_{HI}}{M_{\odot}\,pc^{-2}}\right)},
\end{equation}
where the function $\chi$, in turn, is basically dependent on the metallicity:
\begin{equation}
\chi_{KRU}=3.1\,\frac{1+3.1 Z/Z_{\odot}^{0.365}}{4.1}
\end{equation}
These expressions are valid for $0 < s_{KRU} < 2$ and actually  represent functions depending on metallicity and gas density.
Models using these equations are called KRU.

\begin{figure}
\includegraphics[width=0.45\textwidth,angle=0]{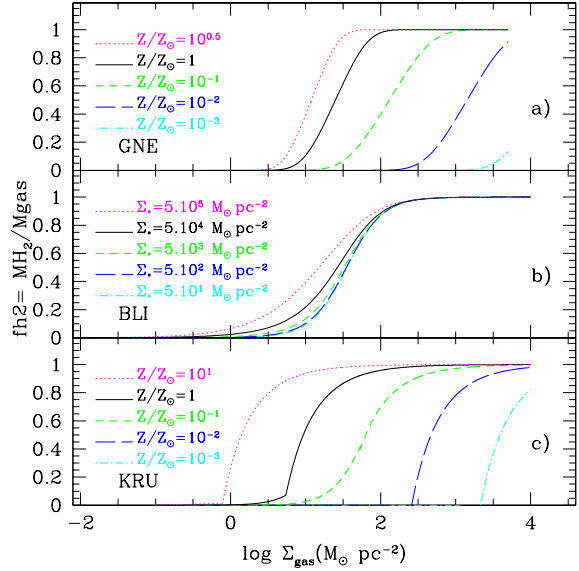}
\caption{The ratio $f_{H_{2}}$ following the prescriptions used in GNE, BLI and KRU models. In Panel a) the prescriptions from GNE  with different lines corresponding to different dust-to-gas ratios  $D_{MWG}$ computed from the metallicity values (see text). In Panel b) the BLI prescriptions which depend on the stellar metallicity, as labelled, defining the pressure in each region: In Panel c) the KRU prescriptions depending mainly on the parameters $\chi_{KRU}$ and $s_{KRU}$, which actually are functions of the metallicity except for the $H_{2}$ low density regime where a dependence on the  total density appears.}
\label{fh2}
\end{figure}

It should be mentioned that in \citet{kru13} the author performs a careful analysis of the formation of stars (and consequently of the molecular cloud formation) in the low  molecular gas regime. According to his work, it is necessary to compute a minimum gas depletion time $\tau_{dep}$ (in Gyr), which the author obtained from his Eqs. (21), (22) and (27), and  then use this value to calculate $f_{H_{2}, KRU}=2/\tau_{dep}$. This technique produces different curves for very low $f_{H_{2}, KRU}$, as appropriate for the early evolution of galaxies, and for the normal regime, when this parameter reaches  values higher than 0.2--0.3 (see his Fig.1 where this behavior is shown). We have also included this method in our code to choose a different value of $f_{H_{2},KRU}$ in the low molecular gas regime. As seen in our Fig.~\ref{fh2}, it also makes the agreement with the GNE prescriptions improve. As in BLI model, the equations of the evolution for both phases of gas are similar to the ones given for GNE, with a similar value of $\kappa_{c,KRU}=f_{H_{2}}/(1-f_{H_{2}})$ in which $f_{H_{2}}$ is calculated as explained.

Panel c) of Fig.\ref{fh2} shows the evolution of $f_{H_{2}}$ with gas surface density for different values of the metallicity. This metallicity, as in the case of the GNE model, varies continuously in our chemical evolution code. However, the prescriptions of GNE and KRU result in rather different  $f_{H_{2}}$ functions, as shown. 

\section{Results}

\subsection{Comparison with the Milky Way Galaxy}
\small
\begin{table*}
\caption{Models computed with different HI to H$_{2}$ transition prescriptions}
\begin{tabular}{llll}
Name & Dependence & Color & Reference\\
\hline
STD & Gas density and  Morf.Type (Eq.~30) & orange &MD05 \\
MOD & Gas density and Morf.Type (Eq.~32) & magenta & This work\\
ASC &  Gas and stellar density and metallicity & red & \citet{asc17}\\
GNE & Gas density, dust,  and FUV Flux   &   blue & \citet{gne11} \\
BLI   & Gas and Stellar  Surface density &  green & \citet{bli06}\\
KRU & Gas density and metallicity & cyan & \citet{kru08,kru09} \\
          &                                              &         & \citet{kru13} \\
\hline
\end{tabular}
\label{models}
\end{table*}
\normalsize
\begin{figure*}
\includegraphics[width=0.7\textwidth,angle=-90]{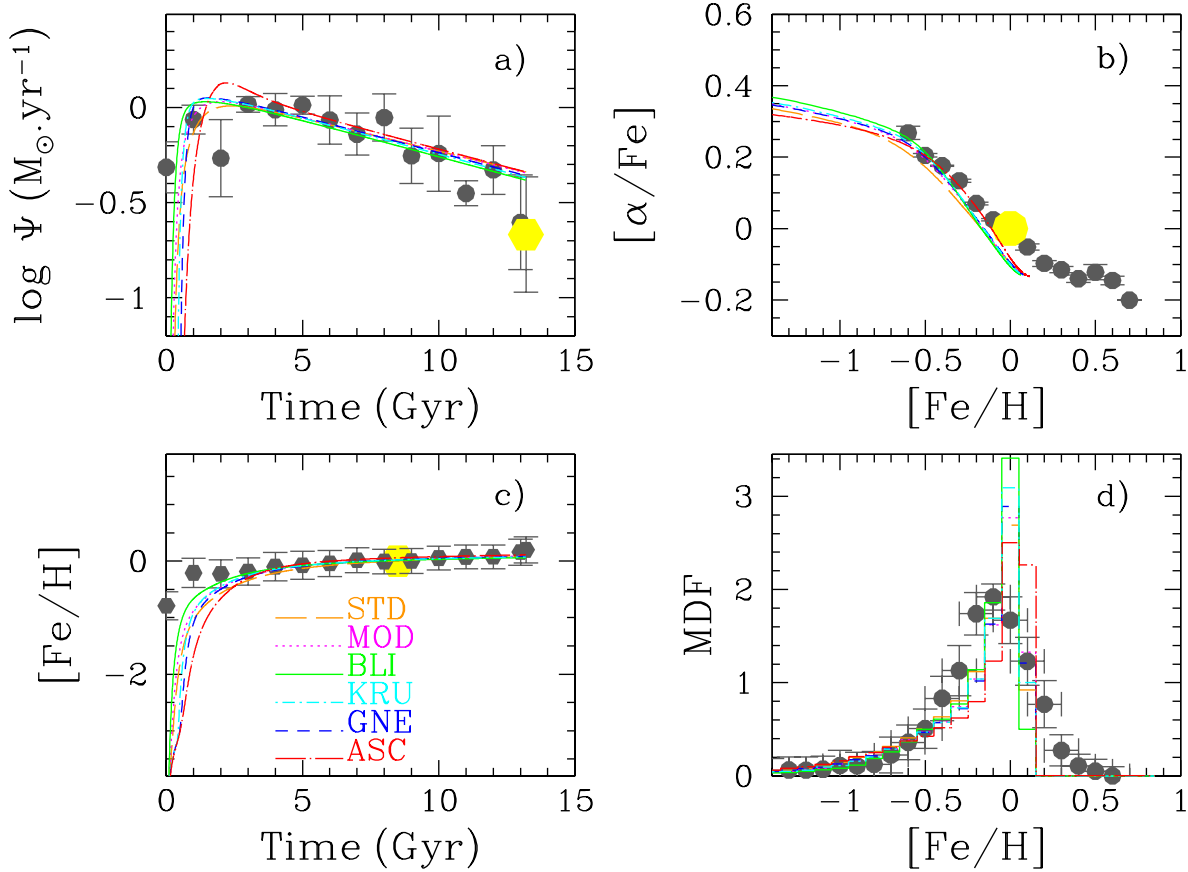}
\caption{The evolution of the Solar Vicinity: a) The SFR history; b) The relative abundance $\rm <[\alpha/Fe]>$  {\sl versus} the iron abundance $\rm [Fe/H]$; c) The enrichment history as Iron abundance $\rm [Fe/H]$ as a function of time; d) The metallicity  distribution function MDF. In all cases we show our six models as labelled in panel c) and data correspond to the observations binned and compiled in \citet[][ Appendix]{molla15}.}
\label{sn}
\end{figure*}

In this subsection we compare the results of {\sc MULCHEM} applied to the MWG using the different prescriptions to form molecular clouds. 
The basic model is the same as described in \citet{molla15} and in the above section 2.  The scenario starts with a protogalaxy with a virial mass of $\sim 10^{12}$\Msun\ assumed spherically distributed following a \citet{burker} profile. This mass, initially in gas phase, falls over the equatorial plane forming out the disc.
The infall rate is assumed as the necessary to create, along a Hubble time, a disk with a radial profile as observed. Details about the evolution
of this infall rate and comparison with existing data and with cosmological simulations results are given \citet{molla16}.  Once this gas in the disk,
molecular clouds are created from the diffuse gas, and then, by cloud-cloud collisions and by the interaction of  massive stars with these clouds, stars 
form.
\begin{figure*}
\includegraphics[width=0.7\textwidth,angle=-90]{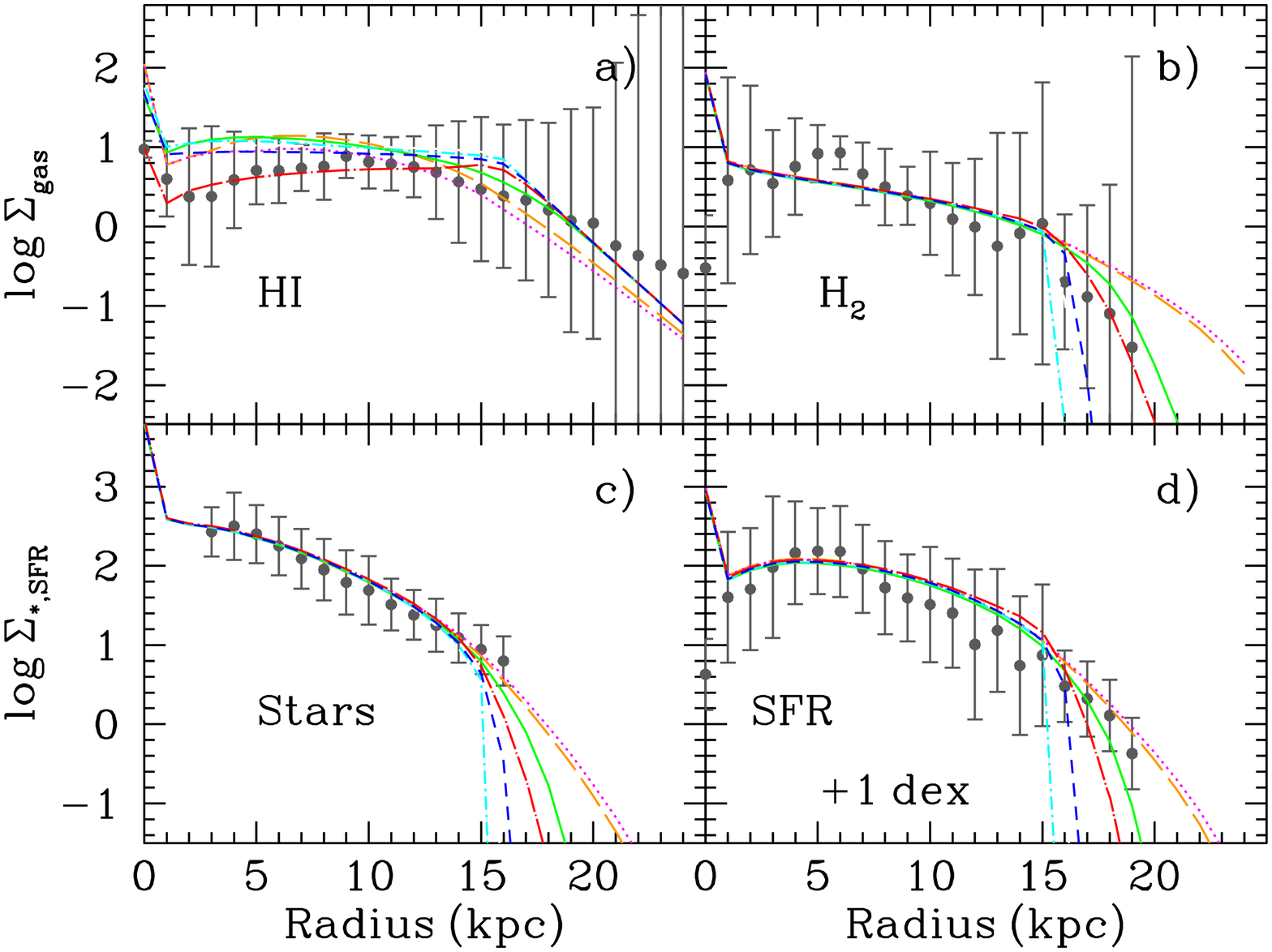}
\caption{The radial distributions of a) HI, b) H$_{2}$, c) stars and d) star formation surface densities.
Models shown with same line coding as Fig.~\ref{sn}, and data correspond to the observations binned and compiled 
in \citet[][Appendix]{molla15}.}
\label{disk}
\end{figure*}
We run 6 models, summarized in Table~\ref{models}. The only differences among the six MWG models are the different methods to convert HI in H$_{2}$ and in the values of 
$\kappa_{c}$, which are computed from the corresponding efficiencies  $\epsilon_{c}$ and volumes in each radial region for STD and MOD models, while  are obtained as explained in the above section for the other four ones. In Table~\ref{models} we give for each model the main characteristics on which the conversion of HI to H$_{2}$ depends and also the color of the line used in our figures for the MWG models and the references in which we base our work. In the first two models, the conversion of diffuse gas, $g$, into molecular gas $c$, occurs by assuming a Schmidt law $\propto g_{D}^{n}$ with a power slope $n=1.5$. In the last 4 models, however, the transition from diffuse to molecular phase is assumed as a linear process defined by $\frac{dc}{dt}=\kappa_{c}\frac{dg}{dt}$. The equations given this ratio as a function of the total density of gas or as a function of other
quantities are given in the previous section for the different authors. 

\begin{figure}
\includegraphics[width=0.48\textwidth,angle=0]{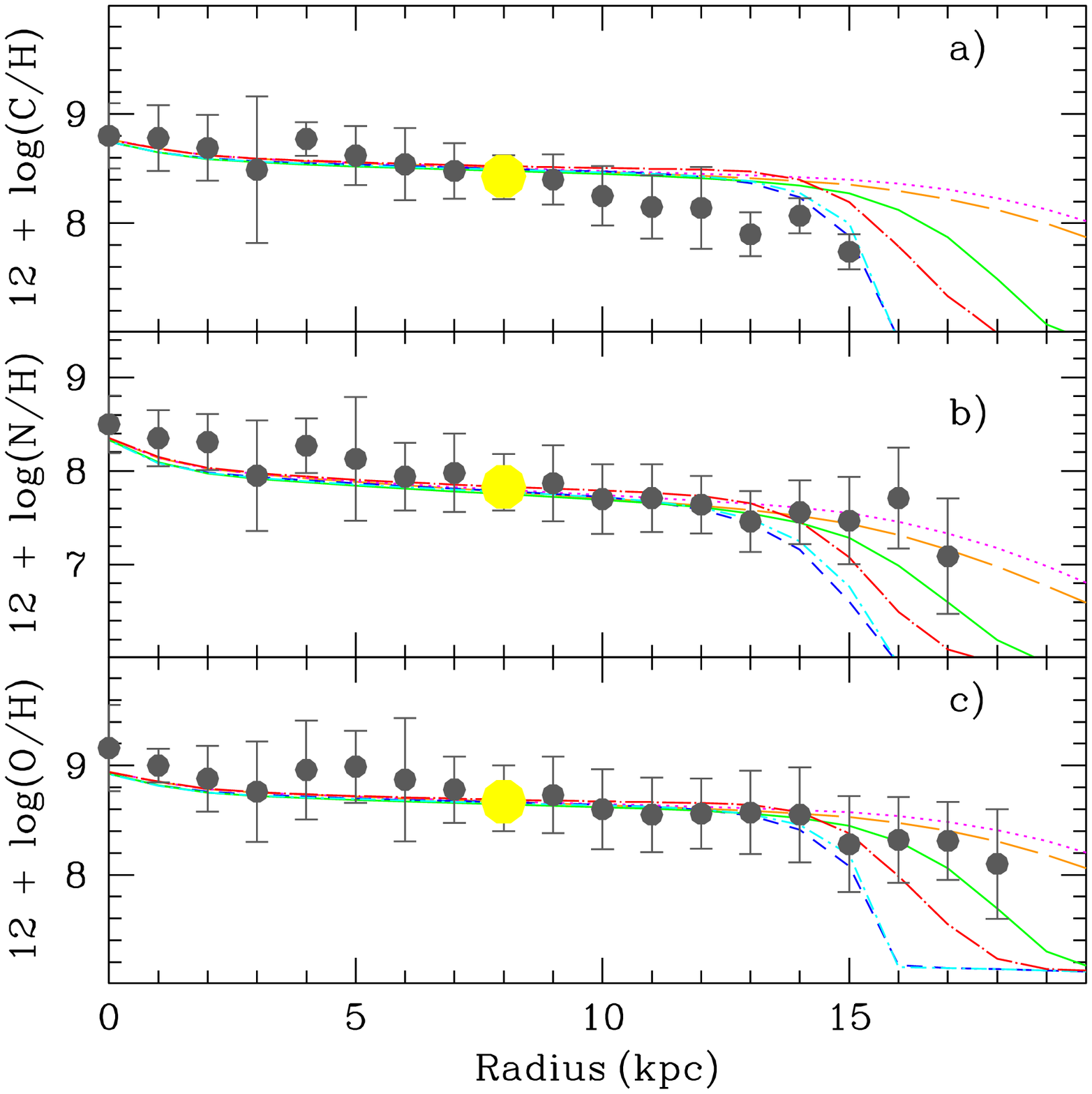}
\caption{The radial distributions of a) C, b) N, and c) O s
Models shown with same line coding as Fig.~\ref{sn}, and data correspond to the observational fits
found in \citet{molla15}.}
\label{abun}
\end{figure}

 In Fig.~\ref{sn} we represent the time evolution of the Solar Region showing the SFR and the enrichment histories as a function of time (left panels) and the alpha-element abundance, [$\rm \alpha$/Fe], and the MDF as a function of the Iron abundance, $\rm [Fe/H]$, taken as a proxy for time (right panels). In panels a), b) and c), it is rather evident that results for the final times, including the present one, are very similar for the six models. However, some differences appear at early times. Thus, all models show a  maximum in the SFR, but for models GNE, BLI, and KRU this maximum is reached  before the other three. In particular, ASC model has this maximum shifted towards slightly later times than the rest of models ($t \sim 2$\,Gyr instead $t<1$\,Gyr) and this maximum is higher.

These differences translate for the enrichment history during the first 3\,~Gyr shown in panel a) into a slower increase in ASC, as expected from a later SFR, while in BLI , the Fe abundance increases very rapidly, as corresponds to the earliest SFR. On the other hand, all the other models, GNE, KRU, STD and MOD behave similarly.  After 4\,Gyr all models show practically the same results, with ASC reaching the highest Fe abundance despite starting to increase later than them all.  Differences among models are, however, smaller than errors. In panels c) and d) differences appear as a consequence of these distinct early SFR, with BLI showing slightly higher [$\alpha$/Fe] at the 
poor-metal end of the correlation, while ASC showing the lowest. Above $\rm [Fe/H]> -1$\,dex, lines cross, ASC being the closest to data at face value.
 The corresponding MDFs are not equal, with ASC showing a larger proportion of metal-rich stars and  all the other models showing a maximum higher than observed around the solar abundance. ASC is the only one that does not show a high peak at solar $\rm [Fe/H]$ as the other models do. However, some of them, as BLI, reproduce better the metallicity distribution in the range $\rm -0.3 < [Fe/H]< 0$.

In Fig.~\ref{disk} we show the corresponding radial distributions of the surface density of both phases of gas, stars and SFR. These radial distributions result very similar
for the six models  within a radius $R\sim 14$\,kpc (just where the optical radius is defined) with all of them showing a good fit to the radial profiles and it is not possible to distinguish the different lines in panel b), c) and d).  The reason for this is that the same total mass is chosen for the six models and the infall rate is also the same for all of them; simultaneously, the SFR is quite similar too. There are, however, important differences out of this galactocentric distance. This occurs because, at variance with our classical models STD and MOD in which the SFR begins at the initial time $t=0$, the other four models require a threshold value for the total or stellar density or the metallicity to initiate the formation of molecular clouds and the subsequent SFR. In fact, we have added a threshold metallicity  $\sim 0.004$ to the Z term in GNE, KRU and ASC models, to avoid the parameter $\kappa_{c}$ to be always zero (and the SFR too) thus preventing any evolution.   This threshold translates into an abrupt increase of $\kappa_{c}$ in the last 4 models when the required conditions are reached and, simultaneously, a strong decrease of the star and gas surface densities in the outer regions of disks in different degrees, due to low densities or metallicities.  Panel a) showing the diffuse gas surface density is where the largest differences arise among the different models. This is expected since they differ in the  prescriptions to convert this diffuse gas into molecular gas. At any rate, all of them show a flatter radial distribution than the other three panels, with ASC showing the best agreement with data up to a galactocentric distance of 14-15\,kpc, while the others look slightly high, in the limit of the error bars. 
 The radial distributions resulting from models KRU, GNE and ASC, on the other hand, show a maximum just at $R=15-16$\,kpc, that we interpret as due to a strong decrease in the parameter $\kappa_{c}$ and that can also be seen in the decreasing of the distributions in panels b), c) and d). This way, outer than 15-16 kpc, BLI shows a better agreement with data than ASC and also better than KRU, and GNE.
  Nevertheless, we should bear in mind that the large errors in the data corresponding to the outer regions do not allow to discriminate among the different models. In panel b),  the comparison is clearer: all models behave basically in the same way  within the optical disc. At face value, BLI and ASC fit better the observations in the outer disk, while  STD and MOD show values higher than expected, and KRU and GNE show a very steep decrease. In any case, it is again necessary to take into account the large error bars of these estimates. In c) all models are plausible up to a galactocentric distance of 14-15\,kpc and differences arise out of this radius. As there are no data out of 16\,kpc any model seems valid.  In panel d) the data points lie between BLI and ASC models and STD and MOD. Summarizing, GNE and KRU yield the worst agreement with data, while BLI and ASC seem to provide the best one.

In Fig.~\ref{abun} we show the radial distribution of C, N and O elemental abundances of the same
six models compared with the data. Again we only see differences in the outer disk regions, as in panels b), c) and d) of the previous Fig.~\ref{disk}. Models GNE, KRU and ASC show a break in the radial gradients while STD, MOD show a smooth behavior with a continuous and slight decrease. BLI is in between the two types. Following O and N data, which arrive until 17-10\,kpc, the classical models STD ad MOD
behave better than models using new prescriptions. However, C data shows a stepper radial gradient, mainly in he outer disk, more in agreement with models GNE, BLI, KRU and ASC. If the radial gradients are steeper or flatter in the outer disk is still a matter of discussion, but there are growing evidences that there exist an universal radial gradient (when measured as function  of a normalized radius) which shows a flattening beyond 2R$_{eff}$ \citep[][ and references therein]{sanchez14,men16}, which would fall around 12--13\,kpc in MWG. If this is the case, the strong decrease of the radial distributions shown in Fig.~\ref{disk} and Fig.~\ref{abun} by models GNE, KRU and ASC would be in total disagreement  with observations, with only BLI as marginally valid, and STD and MOD the only actually able to maintain a certain {\it flattening} although not totally enough.

It is necessary to take into account, when analyzing these results, that all the new prescriptions, that is, GNE, BLI, KRU and even ASC models, have some parameters taken as constant which  actually might be changed, thus modifying the results. Thus, GNE uses a parameter $U_{MWG}$ which we take as 1, just because we are calculating a model for MWG. This parameter, however, could be slightly different along the galactocentric radius,  since it depends basically on the ultraviolet stellar radiation at 1000\,\AA\, which, in turn, would change with the number of massive stars born in each time and region. With a lower value, $U_{MWG}=0.1$, (see Fig.~3), a high value of $f_{H_{2},GNE}$ may be reached at low densities of gas. Therefore, the SFR will be higher, since it depends on the  molecular gas, and also the stars; simultaneously, a strong decrease of the diffuse gas will be produced. On the contrary, with a higher value, as $U_{MWG}=10$, the molecular clouds and stars formation will decrease and, therefore, HI density will maintain high. Using a parameter $U_{MWG}$ variable with radius would do possible to fit the outer regions of the Galactic disk  better than with a constant value. We have checked this idea with  $U_{MWG}=0.5$ in $R=22$\,kpc and a straight line from $R=14\,$kpc and $U_{MWG}=1$. We have seen that effectively the fit improves in the outer disk beyond 14\,kpc. It will be necessary to analyze carefully this subject and its consequences in a future work, after calculating the spectral energy distribution of each time and region, and the corresponding variation of the intensity of 1000\,\AA\, in order to include a more precise dependence on this parameter. Similarly it occurs with other quantities included in the other prescriptions, as the local number density of dust grains in ASC, calculated with a fiducial value $<\sigma\nu>=6\times10^{-17}\,cm^{3}\,s^{-1}$ while other authors use the half of this value \citep{wolfire08}, or the functions $s$ or $\chi$ included in KRU, which are obtained through approximations and simplifications by using typical values of the number density, dust cross section, and H$_{2}$ formation rate coefficient, to compute a value $\chi \sim 1$, which may also be different in other times or other radial regions of the MWG. These considerations do that the proportionality factors included in Eq.(37), (43), (63) and (64), might vary a factor between 2 and 5. This implies that some uncertainties are associated to these model results, and therefore to select the best one is not a straightforward  task.

\subsection{The evolution of the radial distributions}

\begin{figure}
\includegraphics[width=0.48\textwidth,angle=0]{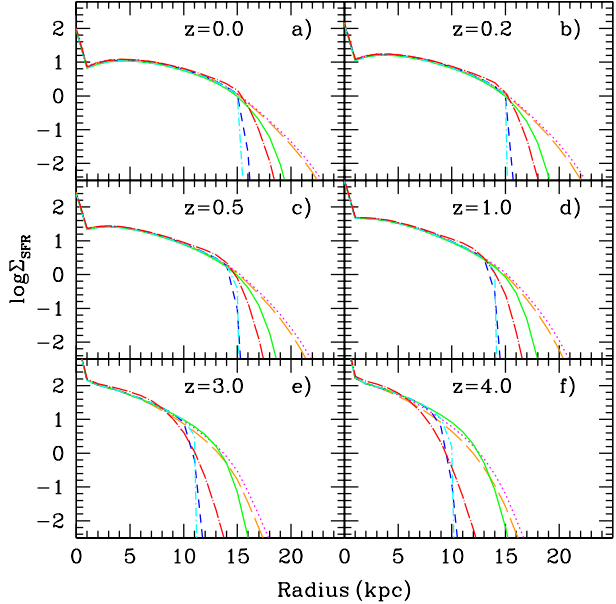}
\caption{The radial distribution of  the SFR surface density as ,$\log{\Sigma_{SFR}}$ for our six models, for a different redshift in each panel as labelled: a) z=0.0; b) z=0.2; c) z=1.0; d) z=3.0; e) z=4.0 and f) z=5.0}
\label{sfr_z}
\end{figure}

\begin{figure}
\includegraphics[width=0.48\textwidth,angle=0]{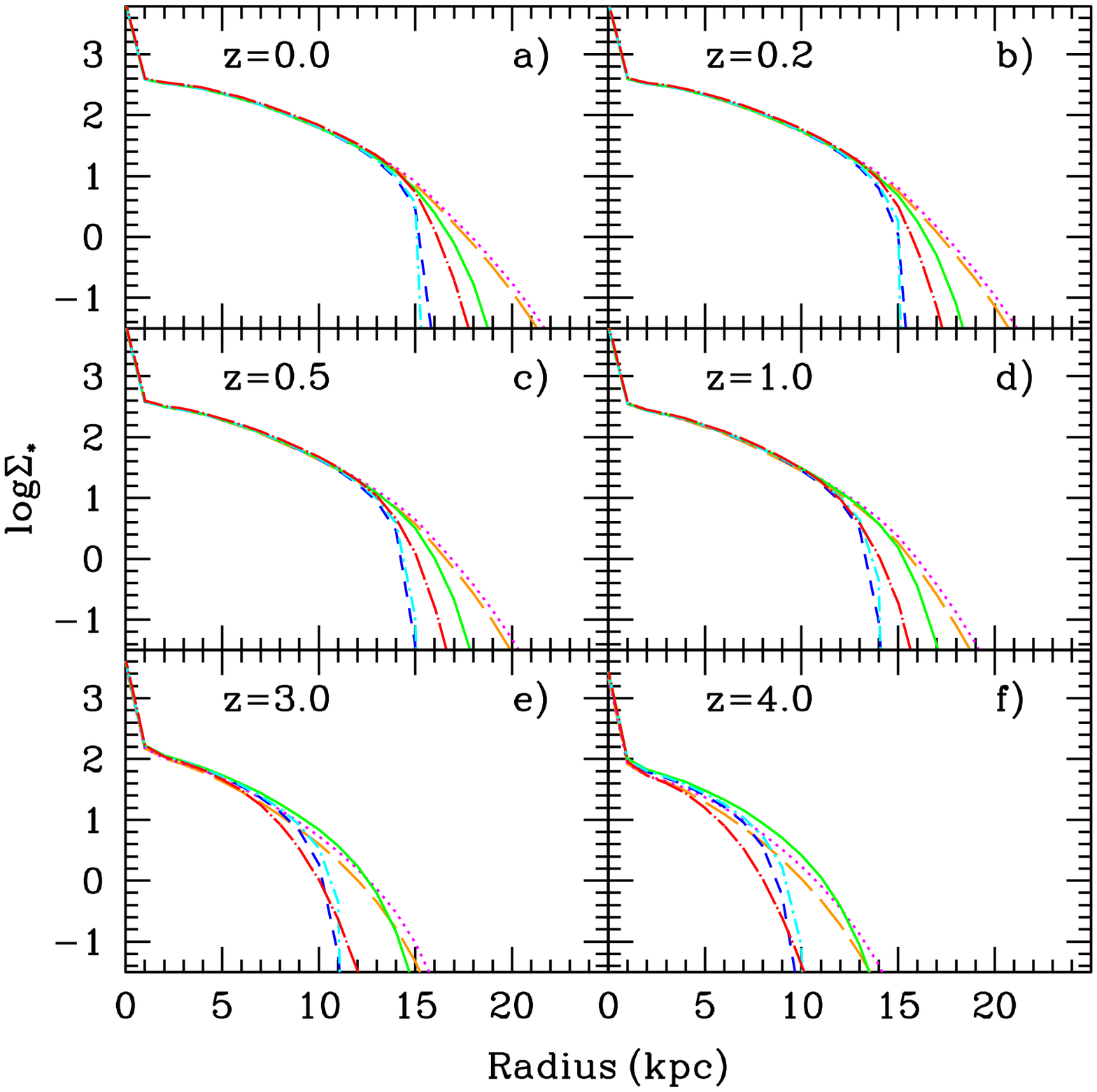}
\caption{The radial distribution of  the stellar surface density, as $\log{\Sigma_{*}}$, for our six models, for a different redshift in each panel as labelled: a) z=0.0; b) z=0.2; c) z=1.0; d) z=3.0; e) z=4.0 and f) z=5.0}
\label{est_z}
\end{figure}

\begin{figure}
\includegraphics[width=0.48\textwidth,angle=0]{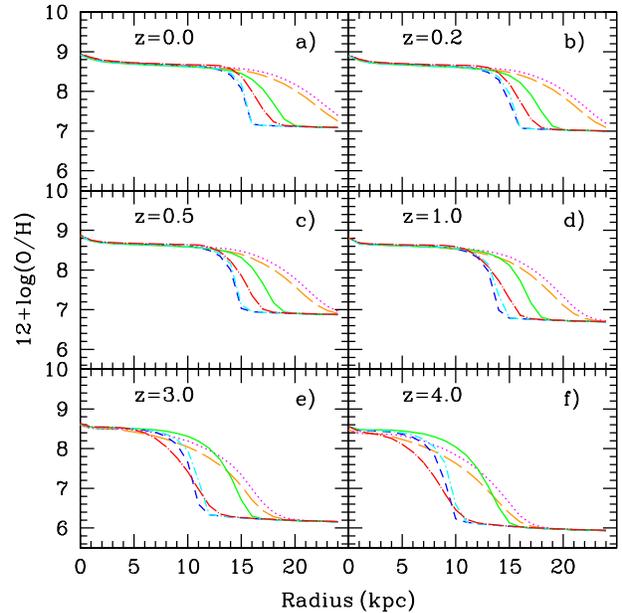}
\caption{The radial distribution of  the oxygen elemental abundance as a function of radius $R$ (in kpc)
for our six models, as $12+\log{(O/H)}$, for a different redshift in each panel as labelled: a) z=0.0; b) z=0.2; c) z=1.0; d) z=3.0; e) z=4.0 and f) z=5.0}
\label{oh_z}
\end{figure}

We plot in Fig.~\ref{sfr_z}, Fig.~\ref{est_z} and Fig.~\ref{oh_z} the radial distributions of star formation rate surface density, stellar profile surface density and the oxygen abundance $12+\log{(O/H)}$, obtained for the six models in different redshifts/evolutionary times.
To transform our evolutionary times, $t$ in redshifts, $z$, we have used the relationship between $z$ and $t$ as given by \citet{mac06}, as explained in \citet{molla14}, using a flat cosmological model with $\Omega_{\Lambda}=0.685$ and $H_{0}=67.3\,km\,s^{-1}\,Mpc^{-1}$ \citep{ade14}. Moreover, we calculate the evolution for 13.2\,Gyr, by assuming an Universe age of 13.8\,Gyr, that is, there is a time shift $\Delta t=0.6\,Gyr$ to start the galaxy formation. We see in Fig.~\ref{sfr_z} to
Fig.~\ref{oh_z} that the models differ mainly at the outer disk, $R> 15 \,kpc$, for $z=0$. However, when we look at the higher redshifts distributions, differences are larger at shorter radius.  In particular, at $z=4$ each radial distribution seem different of the others.  This is particularly clear for the O/H abundance radial distributions. 

\begin{figure}
\includegraphics[width=0.48\textwidth,angle=0]{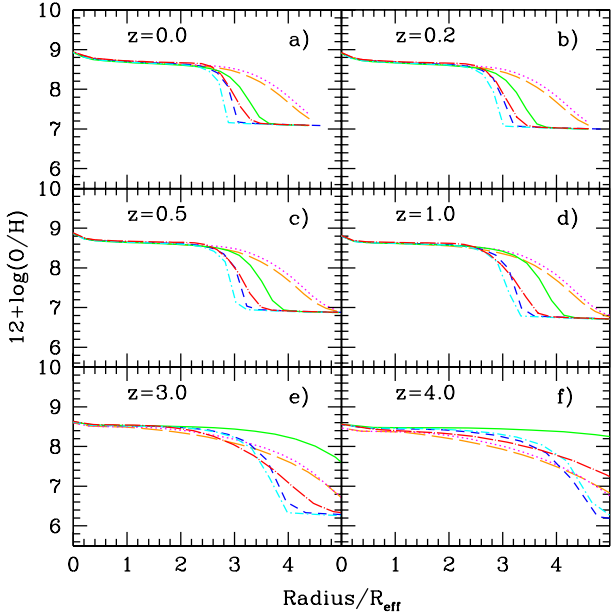}
\caption{The radial distribution of  the oxygen elemental abundance as a function of the normalized radius $R/R_{eff}$, for our six models, as $12+\log{(O/H)}$, for a different redshift in each panel as labelled: a) z=0.0; b) z=0.2; c) z=1.0; d) z=3.0; e) z=4.0 and f) z=5.0}
\label{ohreff_z}
\end{figure}

However, if we represent $12+\log{(O/H)}$ as a function of the normalized radius $R/R_{eff}$, 
Fig~\ref{ohreff_z}, where the effective radius, $R_{eff}$, is defined as the radius which encloses the half stellar mass of the galaxy,  we see that all models coincide until $R=R_{opt}$, where, by definition, $ R_{opt}=3.2\,R_{D}\sim 2.2\,R_{eff}$. This occurs for all redshifts until $z=4$. For redshifts higher that this value it is difficult to estimate the effective radius, which is smaller than 1\,kpc in most of cases. Since we compute our models with a radius step $\Delta\,R=1$ or even 2\, kpc, our results are not longer
accurate enough. 

The fact that the radial gradient be the same when it is measured as a function of a normalized radius related with the stellar population existing in each time is reasonable, since stars are who produce the elements and eject them to the interstellar medium. Therefore,  a similar profile for the region where the stars have been created in each time is expected for all models. Differences must be appears when the star formation is not longer the same: in the outskirts of galaxies when the effect of a threshold, which do not appear equally in all models, is evident.



\begin{figure}
\includegraphics[width=0.35\textwidth,angle=-90]{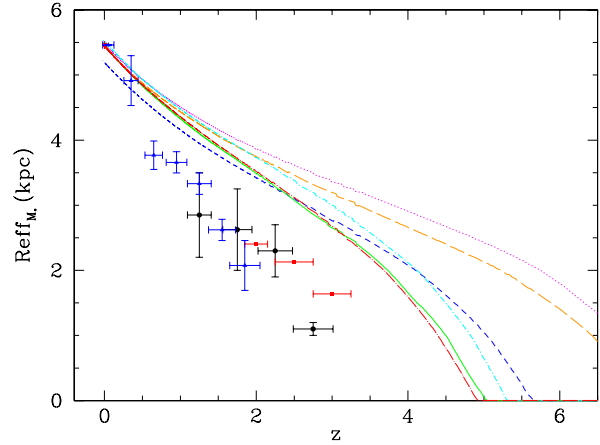}
\caption{The evolution  of the effective radius $R_{eff}$ along the redshift $z$. Models are coding with the same lines as previous figures.
Black full dots and red squares correspond to the observational data given by \citet{tru07} and \citet{bui08}, respectively.}
\label{reff}
\end{figure}

The effective radius, as shown in Fig.~\ref{reff}, is different for each model, showing the fact that the SFR history is not the same for every one, as due to the distinct time scale to create the molecular clouds obtained from the different prescriptions described in Section 2. In this figure, it is evident that each model creates the stellar disk to a different rate, the ASC model being the one that does it more later, compared with the others, and being also the one more in agreement with the observational data from \citet{tru07,bui08}, shown as dots and squares. These authors obtain their data from CANDELS  for galaxies with stellar masses $M_{*} > 10^{10}$\,\Msun. Although they divide their sample in two bins with disks and spheroids, respectively, and we use the ones for the disk galaxies, the effective radius is computed including the bulges, and, moreover, are biased towards massive early galaxies, and maybe for this reason their averaged effective radii are shorter than those we find for a MWG-like model (whose stellar mass is within their data range). The use of the new prescriptions in models GNE, BLI, KRU and ASC produces disks more slowly than the use of a free parameter, as in models STD and MOD in which the formation of the disk occurs very quickly ($z>6$); in GNE and KRU models this happens at $z=5-5.5$ and in ASC and BLI at $z<5$. On the other hand, GNE and KRU cross at $z\sim 4$ and GNE also crosses with two other: ASC and BLI, at $z\sim2.5$. GNE reaches the smallest $R_{eff}$ at $z=0$. BLI and ASC  show a very similar behaviour in this figure.

\subsection{The relation H$_{2}$/HI as function of the radius}

In this section we are going to compare our models with the correlation found by \citet{bigiel08,bigiel12}. In the first work, a correlation between 
both components H$_{2}$ and HI as a function of the normalized radius is found for seven galaxies. Following their findings:
\begin{equation}
\log{\frac{\Sigma_{H_{2}}}{\Sigma_{HI}}}=1.0 -0.977 \times\,R/R_{eff},
\end{equation}

And a similar relationship was also found by \citet{leroy08}, for which
\begin{equation}
\log{\frac{\Sigma_{H_{2}}}{\Sigma_{HI}}}=1.025 -0.789 \times\,R/R_{eff},
\end{equation}

\begin{figure*}
\includegraphics[width=0.35\textwidth,angle=-90]{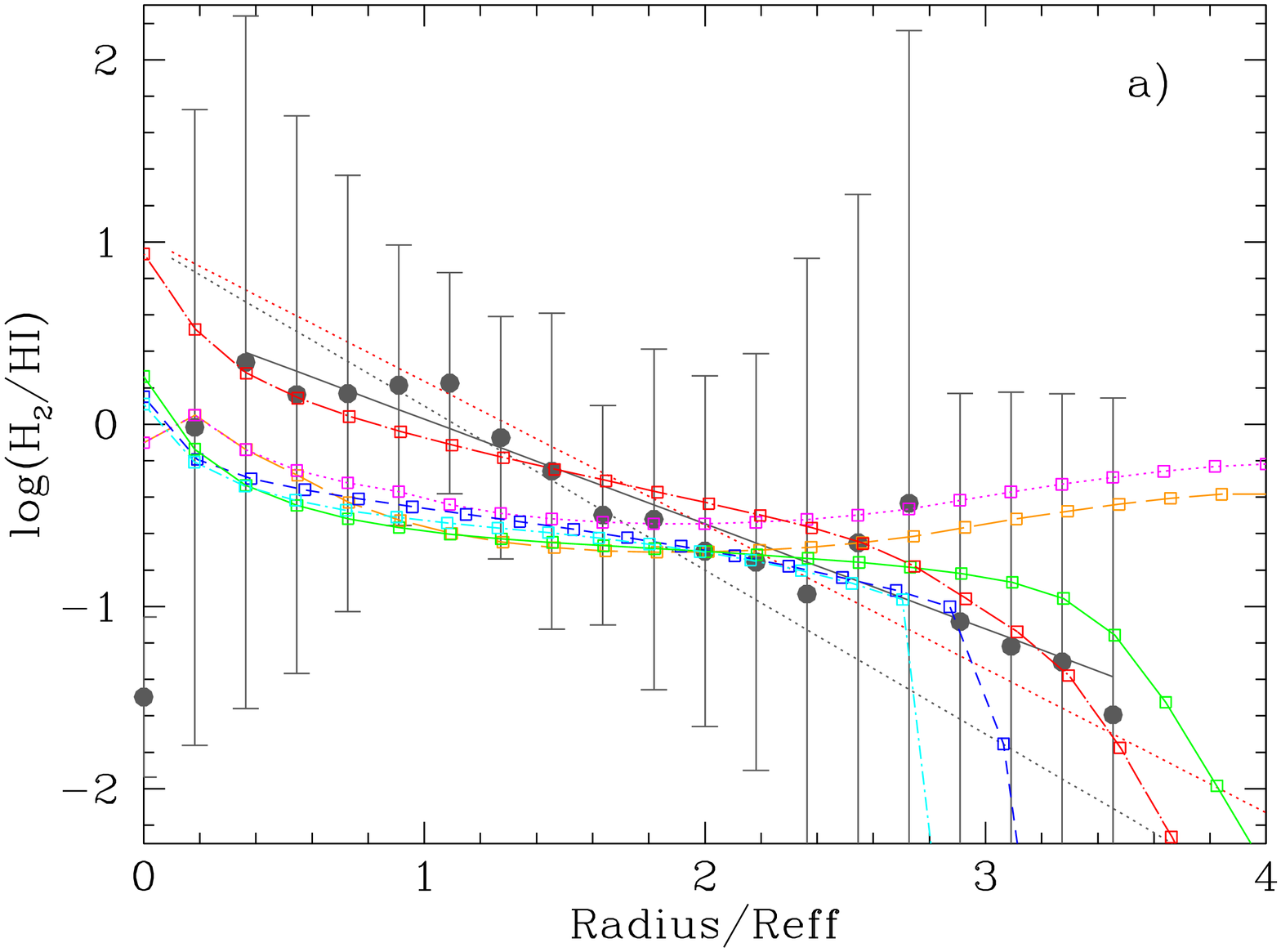}
\includegraphics[width=0.35\textwidth,angle=-90]{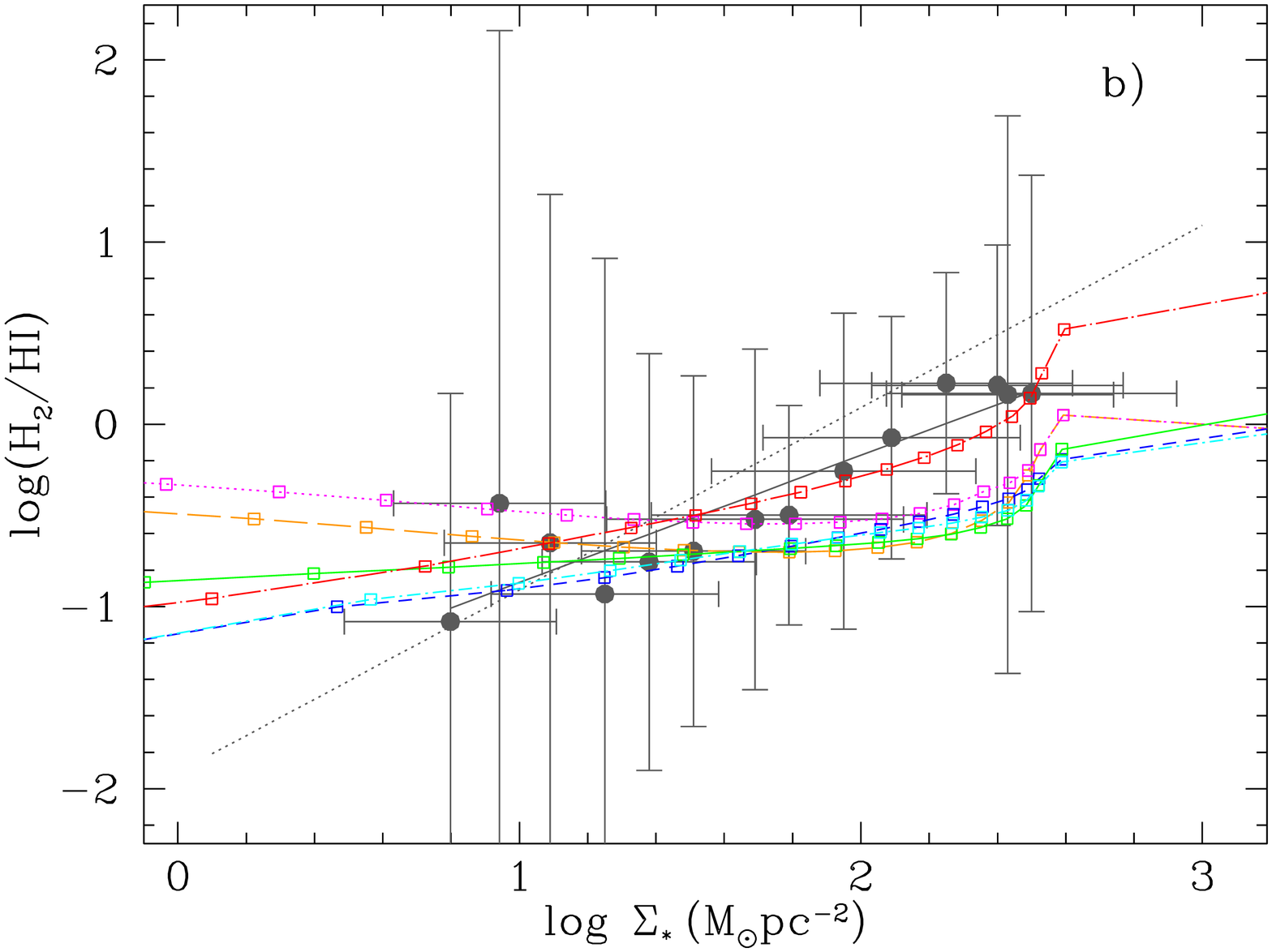}
\caption{a)The radial distribution of the ratio $H_{2}/HI$, in logarithmic scale, for our six models, shown with same line coding as Fig.~\ref{sn}, 
compared with the observational correlation found by \citet{bigiel08} (dotted grey line) and \citet{leroy08} (dotted red line).  b) The relation between the same ratio $\Sigma_{H_{2}}/\Sigma_{HI}$ and the stellar surface density $\Sigma_{*}$, both in logarithmic scale, compared with the relationship from \citet{leroy08} (dotted grey line).  Observational data for MWG galaxy are represented in both panels as back dot with error bars. The solid black line in panel b) is the least squares straight line fitted to these data. }
\label{h2hi}
\end{figure*}

In Fig.~\ref{h2hi}, panel a), we show this radial distribution of the ratio $\Sigma_{H_{2}}/\Sigma_{HI}$, using the normalized radius $R/R_{eff}$, for
the six models compared with the above expressions. In panel b) we show a similar figure where this ratio is represented as a function of the stellar
surface density, compared with the correlation also found by \citet{leroy08}: $\Sigma_{H_{2}}/\Sigma_{HI}=\Sigma_{*}/81\,M_{\odot}\,pc^{-2}$.
In both case we have include the observational data corresponding to MWG as full black dots.

Clearly our old prescriptions as included in STD and MOD models are not as good as the new ones, since they produce very flat radial distributions, even increasing at the
outer regions, in disagreement with the MWG data at face value (although still within errors, which are quite large in these outer regions of the disc). They  also differ
in the observational trends shown in \citet{bigiel08,leroy08}. Models GNE,BLI, and KRU
have a better behavior, but actually only ASC shows a shape closer to the correlations found by these authors.
The same can be seen in the figures at the right panel. At variance with models STD and MOD using the old efficiency to form molecular clouds, 
models computed with the new prescriptions show an increase in the ratio $HI/H_{2}$ with the stellar surface density. However, the slope of these relationships are smaller than the observed one,
ASC being the closest model to that line, although it does not present a straight line in the logarithmic scale but a curve.

\section{Conclusions}

\begin{enumerate}

\item We have computed six different Galactic chemical evolution models in which different prescriptions to form molecular gas from the diffuse gas are
used. 
\item We have checked that the Solar Region and the Galaxy disk data (referring to the radial distributions of diffuse and molecular gas, stellar profile
and star formation rate surface densities) are reproduced as shown in Fig.~\ref{sn}and \ref{disk}. However, it is evident that the ASC prescriptions give 
a model in a better agreement with the data.
\item For what refers to the C, N and O abundances (Fig.~\ref{abun}), all models are able to fit the observations until the optical radius $\sim 13 -15$\,kpc. 
After this radius, in the outer  regions of disk, differences among models arise clearly, with STD and MOD showing more extended disks, while GNE  shows the smallest ones.
ASC, as BLI, shows an intermediate behavior. Given the data for these regions, scarce for C, and reaching far for O, it is quite difficult to determine
which of these models is the best one.
\item We present the evolution along redshift of these quantities obtaining the same conclusion about the size of the disk: STD and MOD show more extended disks
for all redshifts, while ASC and BLI have the smallest for $z>2.5$. However, for $z=0$ GNE has the smallest disk. T
his implies a different grow of the stellar disk in each model, being the ASC model the one which 
starts to increase later, although it arrives to a similar $R_{eff}$ to the other models (except GNE) at $z=0$ (Fig.~\ref{reff}).
\item The Oxygen abundances show a similar slope in their disk distributions when only radial regions within  $\sim 3R_{eff}$ are considered.
It is necessary to take into account that the disk grows at a different rate in each model. In consequence the radial gradient of oxygen abundance
measured within the optical disk is the same for all models and all redshifts, independently of the differences of grow existing among models (Fig.~\ref{ohreff_z}. This point
will need more detailed analysis in a future work.
\item The ratio $HI/H_{2}$ is better reproduced now with the new prescriptions to form molecular clouds from diffuse gas  than before (Fig.~\ref{h2hi}).
This occurs in the  radial distribution of this ratio as a function of the normalized radius $R/R_{eff}$, as in the relationship of this quantity
with the stellar surface density.  
\item This implies that the ASC method to compute the conversion between gas phases, including a dependence on the stellar and gas densities and
on the global metallicity,  seems to produce good results when compared to observational constraints. This model, however, produces a worst fit to the SFR at early times.  Furthermore, the N and O radial gradients seem to be better reproduced by the STD and MOD models at the outer disk.
However, if we want to use a more realistic prescription to form molecular clouds than a free parameter, ASC may be considered the best choice to be used in modern numerical 
chemical evolution models as well as in cosmological simulations.
\end{enumerate}
 
\section{Acknowledgments}
We acknowledge the anonymous referee for very helpful comments which have improved this manuscript.This work has been supported by DGICYT grant AYA2013-47742-C4-4-P and AYA2013-47742-C4-3-P and the European grant SELGIFS FP7-PEOPLE-2013-IRSES-612701.
This work has been supported financially by grant 2012/22236-3 from the S\~{a}o Paulo Research Foundation (FAPESP). This work has made use of the computing facilities of the Laboratory of Astroinformatics (IAG/USP, NAT/Unicsul), whose purchase was made possible by the Brazilian agency FAPESP (grant 2009/54006-4) and the
INCT-A. BKG acknowledges the support of STFC, through grants ST/J001341/1 and ST/G003025/1. MM thanks the kind hospitality and wonderful welcome of the Jeremiah Horrocks Institute at the University of Central Lancashire, the E.A. Milne Centre for Astrophysics at the University of Hull, and the Instituto de Astronomia, Geof\'{\i}sica e Ci\^{e}ncias Atmosf\'{e}ricas in S\~{a}o Paulo (Brazil), where this work was partially done.  The research leading to these results has received funding by the Spanish Ministry of Economy and Competitiveness (MINECO) through the Unidad de Excelencia Mar{\'\i}a de Maeztu CIEMAT - F{\'\i}sica de Part{\'\i}culas grant MDM-2015-0509.


\begin{thebibliography}{}
\bibitem[\protect\citeauthoryear{Planck Collaboration et al.}{2014}]{ade14} 
Planck Collaboration, Ade, P.A.R., et al., 2014, A\&A, 571, 16

\bibitem[\protect\citeauthoryear{Ascasibar et al.}{2015}]{asc15}
Ascasibar, Y., Gavil{\'a}n, M., Pinto, N., et al.\ 2015, MNRAS 448, 2126

\bibitem[\protect\citeauthoryear{Ascasibar et al.}{2017}]{asc17}
Ascasibar,Y., et al. 2017, in preparation

\bibitem[\protect\citeauthoryear{Asplund et al.}{2009}]{asp09} 
Asplund M., Grevesse N., Sauval A.~J., Scott P., \ 2009, ARA\&A, 47, 481 

\bibitem[\protect\citeauthoryear{Bigiel et al.}{2008}]{bigiel08} 
Bigiel, F., Leroy, A., Walter, F., et al.\ 2008, AJ, 136, 2846 

\bibitem[\protect\citeauthoryear{Bigiel \& Blitz}{2012}]{bigiel12} 
Bigiel F., Blitz L., 2012, ApJ, 756, 183  

\bibitem[\protect\citeauthoryear{Blitz \& Rosolowsky}{2006}]{bli06} 
Blitz, L., \& Rosolowsky, E.\ 2006, ApJ, 650, 933 

\bibitem[\protect\citeauthoryear{Buitrago et al.}{2008}]{bui08} 
Buitrago F., Trujillo I., Conselice C.~J., Bouwens R.~J., Dickinson M., Yan H., 2008, ApJ, 687, L61 

\bibitem[\protect\citeauthoryear{Burkert}{1995}]{burker} 
Burkert A., 1995, ApJ, 447, L25

\bibitem[\protect\citeauthoryear{Chieffi \& Limongi}{2004}]{chi04} 
Chieffi A., Limongi M., 2004, ApJ, 608, 405 

\bibitem[\protect\citeauthoryear{Coc et al.}{2012}]{coc12} 
Coc A., Goriely S., Xu Y., Saimpert M., Vangioni E., 2012, ApJ, 744, 158 

\bibitem[\protect\citeauthoryear{Draine \& Bertoldi}{1996}]{draine96}
Draine, B.~T., \& Bertoldi, F.\ 1996, ApJ 468, 269

\bibitem[\protect\citeauthoryear{Eggen, Lynden-Bell, \& Sandage}{1962}]{ELS} 
Eggen O.~J., Lynden-Bell D., Sandage A.~R., 1962, ApJ, 136, 748 

\bibitem[\protect\citeauthoryear{Elmegreen}{1989}]{elm89} 
Elmegreen B.~G., 1989, ApJ, 338, 178 

\bibitem[\protect\citeauthoryear{Ferrini et al.}{1992}]{fer92} 
Ferrini F., Matteucci F., Pardi C., Penco U., 1992, ApJ, 387, 138

\bibitem[\protect\citeauthoryear{Ferrini et al.}{1994}]{fer94} 
Ferrini F., Molla M., Pardi M.~C., Diaz A.~I., 1994, ApJ, 427, 745 

\bibitem[\protect\citeauthoryear{Fu et al.}{2010}]{fu10} 
Fu, J., Guo, Q., Kauffmann, G., \& Krumholz, M.~R.\ 2010, MNRAS, 409, 515 

\bibitem[\protect\citeauthoryear{Gavil{\'a}n, Buell, \& Moll{\'a}}{Gavil{\'a}n et al.}{2005}]{gav05} 
Gavil{\'a}n M., Buell J.~F., Moll{\'a} M., 2005, A\&A, 432, 861 

\bibitem[\protect\citeauthoryear{Gavil{\'a}n, Moll{\'a}, \& Buell}{Gavil{\'a}n et al.}{2006}]{gav06} 
Gavil{\'a}n M., Moll{\'a} M., Buell J.~F., 2006, A\&A, 450, 509 

\bibitem[Glover \& Jappsen(2007)]{glo07}
Glover, S.~C.~O., \& Jappsen, A.-K.\ 2007, ApJ, 666, 1

\bibitem[\protect\citeauthoryear{Gnedin \& Kravtsov}{2011}]{gne11} 
Gnedin N.~Y., Kravtsov A.~V., 2011, ApJ, 728, 88 

\bibitem[\protect\citeauthoryear{Iwamoto et al.}{1999}]{iwa99} 
Iwamoto K., Brachwitz F., Nomoto K., Kishimoto N., Umeda H., Hix W.~R., Thielemann F.-K., 1999, ApJS, 125, 439 

\bibitem[\protect\citeauthoryear{Kroupa}{2001}]{kro01} 
Kroupa P., 2001, MNRAS, 322, 231

\bibitem[\protect\citeauthoryear{Krumholz, McKee \& Tumlinson}{Krumholz et al.}{2008}]{kru08}
Krumhold, M.R., McKee, C.F., \& Tumlinson, J., 2008, ApJ, 689, 865

\bibitem[\protect\citeauthoryear{Krumholz, McKee \& Tumlinson}{Krumholz et al.}{2009}]{kru09}
Krumhold, M.~R., McKee, C.~F., \& Tumlinson, J., 2009, ApJ, 693, 216

\bibitem[\protect\citeauthoryear{Krumholz}{2013}]{kru13} 
Krumholz M.~R., 2013, MNRAS, 436, 2747 

\bibitem[\protect\citeauthoryear{Kubryk, Prantzos, \& Athanassoula}{2015a}]{kub15b} 
Kubryk M., Prantzos N., Athanassoula E., 2015, A\&A, 580, A127 

\bibitem[\protect\citeauthoryear{Kubryk, Prantzos, \& Athanassoula}{2015b}]{kub15a} 
Kubryk M., Prantzos N., Athanassoula E., 2015, A\&A, 580, A126 

\bibitem[\protect\citeauthoryear{Leroy et al.}{2008}]{leroy08} 
Leroy A.~K., Walter F., Brinks E., Bigiel F.,  et al., 2008, AJ, 136, 2782 

\bibitem[\protect\citeauthoryear{Limongi \& Chieffi}{2003}]{lim03} 
Limongi M., Chieffi A., 2003, ApJ, 592, 404 

\bibitem[\protect\citeauthoryear{MacDonald}{2006}]{mac06} 
MacDonald A., 2006, FoPhL, 19, 631 

\bibitem[\protect\citeauthoryear{Moll{\'a}}{2014}]{molla14} 
Moll{\'a} M., 2014, AdAst, 2014, 162949 

\bibitem[\protect\citeauthoryear{Moll{\'a} \& D{\'{\i}}az}{2005}]{md05} 
Moll{\'a} M., D{\'{\i}}az A.~I., 2005, MNRAS, 358, 521

\bibitem[\protect\citeauthoryear{Moll{\'a} et al.}{2015}]{molla15}
Moll{\'a}, M., Cavichia, O., Gavil{\'a}n, M., \& Gibson, B.~K.,\ 2015, MNRAS, 451, 3693 

\bibitem[\protect\citeauthoryear{Moll{\'a} et al.}{2016}]{molla16}
Moll{\'a}, M., D\'{\i}az A.~I., Gibson, B.~K.,  et al., \ 2016, MNRAS, 462, 1329 

\bibitem[\protect\citeauthoryear{Portinari, Chiosi, \& Bressan}{1998}]{pcb98} 
Portinari L., Chiosi C., Bressan A., 1998, A\&A, 334, 505 

\bibitem[\protect\citeauthoryear{Ruiz-Lapuente et al.}{2000}]{rlp00} 
Ruiz-Lapuente P., Blinnikov S., Canal R.,  Mendez J., et al. 2000, MmSAI, 71, 435 

\bibitem[\protect\citeauthoryear{Salucci et al.}{2007}]{sal07} 
Salucci P., Lapi A., Tonini C., Gentile G., et al., 2007, MNRAS, 378, 41 

\bibitem[S{\'a}nchez et al. (2014)]{sanchez14}
S{\'a}nchez, S.~F., Rosales-Ortega, F.~F., Iglesias-P{\'a}ramo, J., et al.\ 2014, A\&A, 563, A49

\bibitem[\protect\citeauthoryear{S{\'a}nchez-Menguiano et al.}{2016}]{men16} 
S{\'a}nchez-Menguiano L., et al., 2016, A\&A, 587, A70 

\bibitem[\protect\citeauthoryear{Sancisi et al.}{2008}]{san08}
Sancisi, R., Fraternali, F., Oosterloo, T., \& van der Hulst , T. A\&ARv, 15, 189

\bibitem[\protect\citeauthoryear{Shankar et al. }{2006}]{sha06} 
Shankar F., Lapi A., Salucci P., De Zotti G., Danese L., 2006, ApJ, 643, 14 

\bibitem[Talbot \& Arnett(1973)]{ta73}
Talbot, R.~J., Jr., \& Arnett, W.~D.\ 1973, ApJ 186, 51

\bibitem[\protect\citeauthoryear{Trujillo et al.}{2007}]{tru07} 
Trujillo I., Conselice C.~J., Bundy K., Cooper M.~C., et al.., 2007, MNRAS, 382, 109 

\bibitem[\protect\citeauthoryear{Young et al.}{1996}]{young96} 
Young J.~S., Allen L., Kenney J.~D.~P., Lesser A.,  et al. 1996, AJ, 112, 1903 

\bibitem[\protect\citeauthoryear{Wolfire et al.}{2008}]{wolfire08} 
Wolfire M.~G., Tielens A.~G.~G.~M., Hollenbach D., Kaufman M.~J., 2008, ApJ, 680, 384-397 


\end{thebibliography}
\end{document}